\documentclass[3p,times,11pt]{elsarticle}
\usepackage[titletoc,title]{appendix}
\usepackage{float}
\usepackage{wrapfig,blindtext}
\usepackage{graphicx} 
\usepackage{epstopdf} 
\usepackage{pslatex} 
\usepackage{amsmath,amssymb}
\usepackage{caption}
\usepackage{subcaption}
\usepackage{amsfonts,amsthm} 
\usepackage[english]{babel} 
\usepackage[dvipsnames]{xcolor}
\usepackage{wasysym}
\usepackage{pdfpages}
\usepackage{float}
\usepackage{comment}

\usepackage{nomencl}

\usepackage{comment}

\newcommand{\MC}{\mathcal}

\usepackage{nomencl}
  \makenomenclature

\journal{Physical Review Fluids}

\begin{document}

\begin{frontmatter}

\title{\sc Non-Markovian Closure Models for Large Eddy Simulations using the Mori-Zwanzig formalism}

\author{Eric J. Parish}\ead{parish@umich.edu}
\author{Karthik Duraisamy}\ead{kdur@umich.edu}
\address{Department of Aerospace Engineering, University of Michigan, Ann Arbor, MI 48109, USA}



\begin{abstract}
 This work uses the Mori-Zwanzig (M-Z) formalism, a concept originating from non-equilibrium statistical mechanics, as a basis for the development of coarse-grained models of turbulence. The mechanics of the generalized Langevin equation (GLE) are considered and insight gained from the orthogonal dynamics equation is used as a starting point for model development. A class of sub-grid models is considered which represent non-local behavior via a finite memory approximation (Stinis, P., ``Mori-Zwanzig reduced models for uncertainty quantification I: Parametric uncertainty," \textit{arXiv:1211.4285}, 2012.), the length of which is determined using a heuristic that is related to the spectral radius of the Jacobian of the resolved variables. The resulting models are intimately tied to the underlying numerical resolution and are capable of approximating non-Markovian effects.  
 Numerical experiments on the Burgers equation demonstrate that the M-Z-based models can accurately predict the temporal evolution of the total kinetic energy and the total dissipation rate at varying mesh resolutions. The trajectory of each resolved mode in phase-space is accurately predicted for cases where the coarse-graining is moderate. LES of homogeneous isotropic turbulence and the Taylor Green Vortex show that the M-Z-based models are able to provide excellent predictions, accurately capturing the sub-grid contribution to energy transfer. Lastly, LES of fully developed channel flow demonstrate the applicability of M-Z-based models to non-decaying problems. It is notable that the form of the closure is not imposed by the modeler, but is rather derived from the mathematics of the coarse-graining, highlighting the potential of M-Z-based techniques to define LES closures.
 \end{abstract}

\end{frontmatter}
\vspace{1cm}

\section{Introduction}
The pursuit of efficient and accurate simulation of turbulent flows continues to present great challenges to the scientific computing community. The continuous cascade of scales makes Direct Numerical Simulation (DNS) methodologies prohibitively expensive for most practical flows. Generally, the solution of high Reynolds number turbulent flows is made tractable by developing a set of surrogate equations that display a reduced range of scales. The reduction of scales  in Large Eddy Simulations (LES) can be viewed as a type of coarse-graining (CG) technique for the Navier-Stokes equations. Typically associated with atomistic simulations, the coarse-graining approach removes the degrees of freedom associated with the microscopic scales and attempts to only compute the macroscopic dynamics~\cite{CGMethods}. The effects of the microscopic scales on the macroscopic scales are expressed through a constitutive relation. This relation is central to a coarse-grained model~\cite{PrinciplesOfMultiscaleModeling}. In a continuum solution of a turbulent flow, the macroscopic processes are the large scale energy-containing flow structures while the microscopic processes can be taken to be the small scale turbulent fluctuations that occur at the Kolmogorov scales.

The prediction of macroscopic quantities in the absence of explicit microscopic information constitutes a classical problem of
multi-scale modeling. For systems that display some degree of separation between the large and small scales, elegant mathematical treatments have been established and significant progress has been made~\cite{OrtizHierarchical}. Systems that exhibit a continuous cascade of scales present a greater challenge as the coarse-graining process leads to non-Markovian effects that are challenging to understand and model. In turbulent flows, the sub-grid effects are generally accounted for through a sub-grid model. The majority of sub-grid scale models for LES are Markovian and are based on scale invariance, homogeneity, and rapid equilibration of small scales. These models are inadequate in many problems and warrant improvement.

The Mori-Zwanzig (M-Z) formalism provides a mathematical procedure for the development of coarse-grained models of systems that lack scale separation. Originating from irreversible statistical mechanics, the M-Z formalism provides a method for re-casting a dynamical system into an equivalent, lower-dimensional system. In this reduced system, which is commonly referred to as the generalized Langevin equation (GLE), the effect of the microscopic scales on the macroscopic scales appears as a convolution integral (which is sometimes referred to as memory) and a noise term. The appearance of the memory term in the GLE demonstrates that, in a general setting, the coarse-graining procedure leads to non-local memory effects. The M-Z formalism alone does not lead to a reduction in computational complexity as it requires the solution of the orthogonal (unresolved) dynamics equation. In the past decade, however, the M-Z formalism has gained attention in the setting of model order reduction. The optimal prediction framework developed by the Chorin group~\cite{ProblemReduction,Chorin_predictwithmem,StatisticalMechanics,ChorinOptimalPrediction,BarberThesis} uses the GLE to obtain a set of equations that describe the evolution of the macroscopic scales conditioned on the knowledge of the microscopic scales. The framework additionally begins to address how the memory convolution integral can be approximated without solving the orthogonal dynamics.

Constructing an appropriate surrogate to the memory integral requires an understanding of the structure of the orthogonal dynamics and its impact on the convolution integral. Obtaining insight into this structure is challenging as it requires the solution of the orthogonal dynamics equation, which is a high-dimensional partial differential equation. Givon et al.~\cite{GivonOrthogonal} prove the existence of classical solutions to the orthogonal dynamics equation for a simple set of projection operators and show the existence of weak solutions in the general case. Hald et al.~\cite{HaldPredictFromData} demonstrate that the memory consists of convolving a sum of temporal covariances of functions in the noise subspace. Chorin et al.~\cite{Chorin_predictwithmem} make use of the fact that the Hermite polynomials are orthogonal with respect to the conditional expectation inner product and develop a set of Volterra integral equations that can be used to approximate the memory integrand. This finite-rank projection is shown to provide a reasonably accurate representation of the memory kernel for a system of two oscillators with a non-linear coupling, but the process is intractable for high-dimensional problems unless a low-dimensional basis is used. Bernstein~\cite{bernsteinBurgers} applies this methodology and uses a first order basis to attempt to gain information about the memory kernel for Burgers equation. The simplicity of the basis functions, however, limits the insight gained from this process.

Despite the challenges in understanding the structure of the memory, various surrogate models exist. The most common approximation to the memory term is the t-model; so named because  time appears explicitly in the closure. The t-model results from both a long memory and a short time approximation~\cite{bernsteinBurgers} and has been applied with varying success to numerous systems. Bernstein~\cite{bernsteinBurgers} applied the t-model to Fourier-Galerkin solutions of Burgers equation. Numerical experiments showed that the t-model accurately characterized the decay of energy resulting from the formation of shocks. Hald and Stinis~\cite{stinisEuler} applied the t-model to two and three-dimensional simulations of the Euler equations. The t-model correctly preserved energy for two-dimensional simulations and provided results consistent with theory for three-dimensional simulations of the Taylor-Green Vortex. Chandy and Frankel~\cite{ChandyFrankelLES} used the t-model in Large Eddy Simulations of homogeneous turbulence and the Taylor-Green vortex. The model was found to be in good agreement with Direct Numerical Simulation and experimental data for low Reynolds number cases, but discrepancies were seen for higher Reynolds numbers. Additionally, the Large Eddy Simulations of the Taylor-Green vortex were performed with high enough resolution that simulations not utilizing any sub-grid model are more accurate than Smagorinsky-type models. Stinis~\cite{Stinis-rMZ,RenormalizedMZ} introduced a set of renormalized models based on expansions of the memory integrand. These models were applied to  Burgers equation and were found to be in good agreement with the full order model. For all of these problems, the reason for the relative success of the t-model has remained a mystery and is an outstanding question in the literature.

Stinis~\cite{stinisHighOrderEuler,stinis_finitememory} proposed a class of approximations to the memory integral that utilize the concept of a finite memory. The models involve the integration of additional differential equations that describe the evolution of the memory integral. The models were applied to Burgers equation and the 3D Euler equations. In these numerical experiments it was found that a finite memory length was required for stability, but the validity of the finite memory assumption was not addressed. This class of models appears to be more capable and robust than the t-model and will be a main consideration of this work. 

While we have restricted our discussion to fluid-dynamics applications, a large body of research regarding the Mori-Zwanzig formalism exists in the molecular dynamics community, for instance, Refs.~\citenum{Darve_CGMZ,Karniadakis_CGMZ}. Additionally, the Mori-Zwanzig approach can be used in contexts outside of coarse-graining. Uncertainty quantification, for example, is one such field where the Mori-Zwanzig formalism has attracted recent attention~\cite{MZ_for_UQ,stinis_UQ2}.

The objective of this work is to extend the applicability of Mori-Zwanzig-based closures to Fourier-Galerkin simulations of turbulent flows and to rigorously investigate their performance in a number of problems. Emphasis will be placed on Stinis' finite memory models~\cite{stinisHighOrderEuler,stinis_finitememory}. This class of closures has not gained substantial exposure in the fluids community. The organization of this paper will be as follows: Section 2 will 
present an introduction to the Mori-Zwanzig formalism. The mechanics of the convolution memory integral and construction of surrogate models will be discussed. In Section 3, the models will be applied to the viscous Burgers equation. In Sections  4 and 5, the models will be applied to the incompressible Navier-Stokes equations, where
homogeneous isotropic turbulence, the Taylor Green vortex, and fully developed channel flow are considered.  In Section 6, conclusions and perspectives will be provided.

\section{Mori-Zwanzig Formalism}\label{sec:M-Z-Formalism}
A brief description of the Mori-Zwanzig formalism is provided in this section. A demonstrative example is first provided to introduce the unfamiliar reader to the Mori-Zwanzig formalism. Consider a two-state linear system given by
\begin{equation}\label{eq:linear1}
\frac{dx}{dt} = A_{11}x + A_{12} y
\end{equation}
\begin{equation}\label{eq:linear2}
\frac{dy}{dt} = A_{21}x + A_{22} y.
\end{equation}
Suppose that one wants to created a `reduced-order' model of the system given in Eqns.~\ref{eq:linear1} and~\ref{eq:linear2} by creating a surrogate system that depends only on $x(t)$. For example,
\begin{equation}\label{eq:linear3}
\frac{dx}{dt} = A_{11}x + F(x).
\end{equation}
The challenge of the numerical modeler is to then construct the function $F(x)$ that accurately represents the effect of the unresolved variable $y$ on the resolved variable $x$. For this simple linear system, $F(x)$ can be exactly determined by solving Eq.~\ref{eq:linear2} for $y(t)$ in terms of a general $x(t)$. Through this process, the two-component Markovian system can be cast as a one-component non-Markovian system that has the form
\begin{equation}\label{eq:linear4}
\frac{dx}{dt} = A_{11}x + A_{12}A_{21}\int_0^t x(t-s)e^{A_{22}s}ds + A_{21}y(0)e^{A_{22}t}.
\end{equation}
Equation~\ref{eq:linear4} has no dependence on $y(t)$ and hence is closed.
This reduction of a Markovian set of equations to a lower-dimensional, non-Markovian set of equations is the essence of the Mori-Zwanzig formalism. The formalism provides a framework to perform this reduction for systems of non-linear differential equations. 

A formal presentation of the Mori-Zwanzig formalism is now provided. The discussion here is adapted from Refs.~\citenum{Chorin_predictwithmem,StatisticalMechanics}. 
Consider the semi-discrete non-linear ODE
\begin{equation}\label{eq:baseODE}
\frac{d \phi }{dt} = R(\phi),
\end{equation}
where $\phi = \{\hat{\phi},\tilde{\phi}\}$, with $\hat{\phi} \in \mathbb{R}^N$ being the relevant or resolved modes, and  $\tilde{\phi} \in \mathbb{R}^M$ being the unresolved modes. The initial condition is $\phi(0) = \phi_0$. 
The non-linear ODE can be posed as a linear partial differential equation by casting it in the Liouville form,
\begin{equation}\label{eq:LiouvilleForm}
\frac{\partial}{\partial t} u(\phi_0,t) = \mathcal{L}u(\phi_0,t),
\end{equation}
with $u(\phi_0,0) = g(\phi(\phi_0,0))$. The Liouville operator is defined as 
$$\MC{L} = \sum_{k=1}^{N+M} R_k(\phi_0) \frac{\partial}{\partial \phi_{0k}},$$
where $\phi_{0k} = \phi_k(0)$.
It can be shown that the solution to Eq.~\ref{eq:LiouvilleForm} is given by
\begin{equation}\label{eq:LiouvilleSol}
u(\phi_0,t) = g(\phi(\phi_0,t)).
\end{equation}
The semigroup notation is now used, i.e. $u(\phi_0,t) =  g(e^{t\MC{L}}\phi_0$). One can show that the evolution operator commutes with the Liouville operator, i.e., $e^{t \MC{L}}\MC{L} = \MC{L}e^{t \MC{L}}$.
Consider the initial conditions to be random variables drawn from some probability distribution $\text{P}(\phi_0)$. Given this distribution, the expected value of a function $g(\phi_0)$ is given by
$$\text{E}[g(\phi_0)] = \int_{\Gamma} g(\phi_0) \rho(\phi_0) d\phi_0,$$
where $\rho$ is the probability density. Assume $\phi \in \Gamma$, where $\Gamma$ is an $L^2$ Hilbert space endowed with an inner product 
$(f,g) = \text{E}[fg]$.
Consider now the coarse-grained simulation of Eq.~\ref{eq:baseODE}, where the variables in $\hat{\phi}$ are resolved and the variables in $\tilde{\phi}$ are unresolved. By taking $g(\phi_0) = \phi_{0j}$, an equation for the trajectory of a resolved variable can be written as
\begin{equation}\label{eq:LiouvilleFormSG2}
\frac{\partial}{\partial t} e^{t \MC{L}}\phi_{0j}
= e^{t \MC{L}}\mathcal{L}\phi_{0j}.
\end{equation}
The term on the right hand side is a function of both the resolved and unresolved variables. To proceed, define the space of  the resolved variables by $\hat{L}^2$. Further, define $\mathcal{P}: L^2 \rightarrow \hat{L}^2$, as well as $\mathcal{Q} = I - \mathcal{P}$. An example of a possible projection operator would be, for a function $f(\hat{\phi}_0,\tilde{\phi}_0)$, $\mathcal{P}f(\hat{\phi}_0,\tilde{\phi}_0) = f(\hat{\phi}_0,0)$. Using the identity $I = \MC{P + Q}$, the right hand side of Eq.~\ref{eq:LiouvilleFormSG2} can be split into a component that depends on the resolved variables and one that depends on the unresolved variables,
\begin{equation}\label{eq:LiouvilleFormSG3}
\frac{\partial}{\partial t} e^{t \MC{L}}\phi_{0j}
= e^{t \MC{L}}\mathcal{PL}\phi_{0j} + e^{t \MC{L}}\mathcal{QL}\phi_{0j}.
\end{equation}
At this point the Duhamel formula is utilized,
$$e^{t \mathcal{L}} = e^{t \mathcal{Q} \mathcal{L}} + \int_0^t e^{(t - s)\mathcal{L}} \mathcal{P}\mathcal{L} e^{s \mathcal{Q} \mathcal{L}} ds.$$
Inserting the Duhamel formula into Eq.~\ref{eq:LiouvilleFormSG2}, the generalized Langevin equation is obtained,
\begin{equation}\label{eq:M-Z_Identity}
\frac{\partial}{\partial t} e^{tL}\phi_{0j} =   \underbrace{e^{tL}\MC{PL}\phi_{0j}}_{\text{Markovian}} +   \underbrace{e^{tQL}\MC{QL}\phi_{0j}}_{\text{Noise}} + 
 \underbrace{ \int_0^t e^{{(t - s)}\mathcal{L}} \mathcal{P}\mathcal{L} e^{s \mathcal{Q} \mathcal{L}} \MC{QL}\phi_{0j}ds}_{\text{Memory}}.
\end{equation}
Equation~\ref{eq:M-Z_Identity} is the Mori-Zwanzig identity. The system described in Eq.~\ref{eq:M-Z_Identity} is exact and is an alternative way of expressing the original system. Equation~\ref{eq:M-Z_Identity} makes a profound statement: coarse-graining leads to memory effects. Note that the convolution integral represents numerical memory, as opposed to physical memory. One can expect the time-scales of this numerical memory to depend both on the physics of the problem at hand and the level of coarse-graining.

For notational purposes, define
\begin{equation}\label{eq:orthoNotation}
F_j(\phi_0,t) = e^{t\MC{QL}}\MC{QL}\phi_{0j} \qquad K_j(\phi_0,t) = \MC{PL}F_j(\phi_0,t).
\end{equation}
By definition, $F_j(\phi_0,t)$ satisfies
\begin{equation}\label{eq:orthogonalDynamicsEq}
\frac{\partial}{\partial t} F_j(\phi_0,t) = \MC{QL}F_j(\phi_0,t),
\end{equation}
where $F_j(\phi_0,0) = \MC{QL}\phi_{0j}$. Equation~\ref{eq:orthogonalDynamicsEq} is referred to as the orthogonal dynamics equation. It can be shown that solutions to the orthogonal dynamics equation live in the null space of $\MC{P}$ for all time, meaning $\MC{P}F_j(\phi_0,t) = 0$. Using the notation in Eq.~\ref{eq:orthoNotation},  
Eq.~\ref{eq:M-Z_Identity} can be written as
\begin{equation}\label{eq:M-Z_Identity2}
\frac{\partial}{\partial t} \phi_j(\phi_0,t) =  R_j(\hat{\phi}(\phi_0,t)) +  
F_j(\phi_0,t) + 
 \int_0^t K_j \big(\hat{\phi}(\phi_0,t-s),s \big)ds.
\end{equation}
A simplification comes from projecting Eq.~\ref{eq:M-Z_Identity2} to eliminate the dependence on the noise term,
\begin{equation}\label{eq:M-Z_Identity3}
\frac{\partial}{\partial t} \MC{P}\phi_j(\phi_0,t) =  \MC{P} R_j(\hat{\phi}(\phi_0,t)) +  
 \MC{P} \int_0^t K_j \big(\hat{\phi}(\phi_0,t-s),s \big)ds.
\end{equation}
Eq.~\ref{eq:M-Z_Identity3} provides a set of equations for $\MC{P}\phi(\phi_0,t)$, the best possible approximation for $\hat{\phi}_j$ on $\hat{L}^2$ given knowledge about the initial density of $\tilde{\phi}$. The evaluation of the memory kernel is, in general, not tractable as it requires the solution of the orthogonal dynamics equation. It does, however, provide a starting point to derive closure models. Additionally, note that the final projection does not necessarily imply a reduction in computational complexity. Although Eq.~\ref{eq:M-Z_Identity3} has no dependence on the noise term, one is left with the challenge of projecting the potentially non-linear Markovian term. For general non-linear functions, the projection operator does not commute ($\text{E}[f(x)] \ne f (\text{E}[x])$).

\subsection{The Orthogonal Dynamics Equation and the Memory Kernel}\label{sec:ortho}
The challenge of the M-Z procedure for model order reduction is to construct an accurate, computationally tractable approximation to the memory integral. The construction of such an approximation requires an understanding of the form of the integrand and the underlying mechanics of the memory kernel. In its primitive form, the orthogonal dynamics equation (Eq.~\ref{eq:orthogonalDynamicsEq}) is a partial differential equation that has dimension $N+M$ (with $N$ being the number of resolved variables and $M$ being the number of unresolved variables). Pursuing solutions to the orthogonal dynamics in this form is not tractable. Recall, however, that the generalized Langevin equation itself is a PDE with dimension $N+M$, but its solution may be obtained by solving a set of ordinary differential equations. It can be shown that, when the original dynamical system is linear, that solutions to the orthogonal dynamics equation can be obtained by solving a corresponding set of auxiliary ordinary differential equations. This methodology was suggested in~\cite{ParishAIAA2016} as a procedure to approximate the orthogonal dynamics in non-linear systems and was further developed and investigated in detail in~\cite{GouasmiMZ1}. To gain insight into the orthogonal dynamics, we consider a simple linear system.

Consider a linear system governed by
\begin{equation}\label{eq:linearOrtho1}
\frac{d\phi}{dt} = \mathbf{A} \mathbf{\phi}
\end{equation}
with $\mathbf{\phi}(0) = \mathbf{\phi}_0$.
The linear system can be split into resolved and unresolved components,
\begin{equation}\label{eq:linearOrtho2}
\frac{d\phi}{dt} = \mathbf{A}^{\MC{P}} \hat{\mathbf{\phi}} + \mathbf{A}^{\MC{Q}} \tilde{\mathbf{\phi}}.
\end{equation}
It can be shown that solutions to the orthogonal dynamics equation can be obtained by solving the auxiliary system
\begin{equation}\label{eq:linearOrtho3}
\frac{d\phi^{\MC{Q}}}{dt} = \mathbf{A}^{\MC{Q}} {\mathbf{\phi}^{\MC{Q}}}
\end{equation}
with $\mathbf{\phi^{Q}}(0) = \mathbf{\phi_0}$. 
Assuming $\mathbf{A}^{\MC{Q}}$ to be diagonalizable, the linear system has the solution
\begin{equation}\label{eq:linearOrthoSolution}
\mathbf{\phi}^{\MC{Q}}(t) =\mathbf{S} e^{\Lambda t} \mathbf{S^{-1} \phi}_0,
\end{equation}
where ${\Lambda}$ and $\mathbf{S}$ are the eigenvalues and eigenvectors of $\mathbf{A}^{\MC{Q}}$. The memory is then given by
\begin{equation}\label{eq:IntegralSolution2}
 \int_0^t e^{{(t - s)}\mathcal{L}} \mathcal{P}\mathcal{L} e^{s \mathcal{Q} \mathcal{L}} \MC{QL}\phi_{0j}ds =  \int_0^t  e^{(t - s) \MC{L}} \MC{PL}   \mathbf{S} e^{\Lambda s} \mathbf{S^{-1}}\MC{QL}\mathbf{ \phi}_0 ds.
\end{equation}
 For cases where the eigenvectors and eigenvalues can be obtained analytically, or are independent of the initial conditions, the memory kernel can be directly evaluated. 

\begin{figure}
\begin{center}
\begin{subfigure}[t]{0.32\textwidth}
\includegraphics[width=1.0\linewidth]{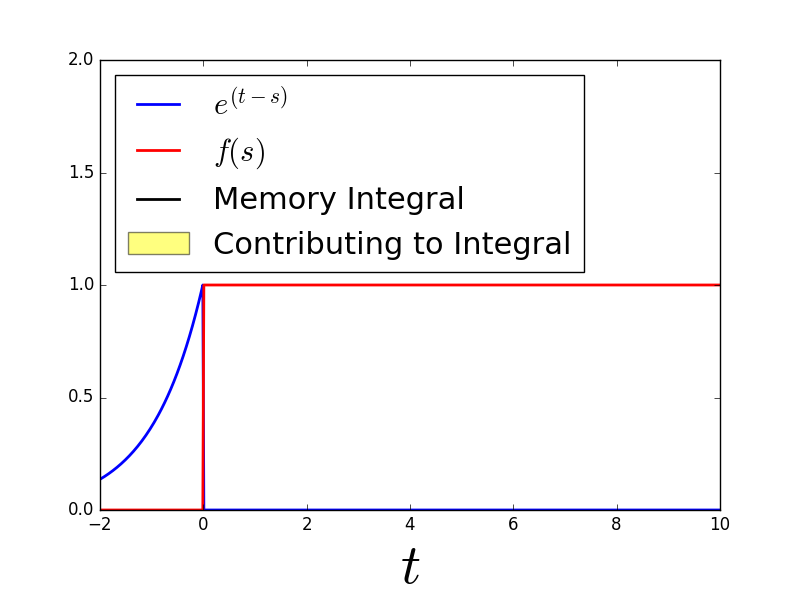}
\caption{$t = 0$.}
\end{subfigure}
\begin{subfigure}[t]{0.32\textwidth}
\includegraphics[width=1.0\linewidth]{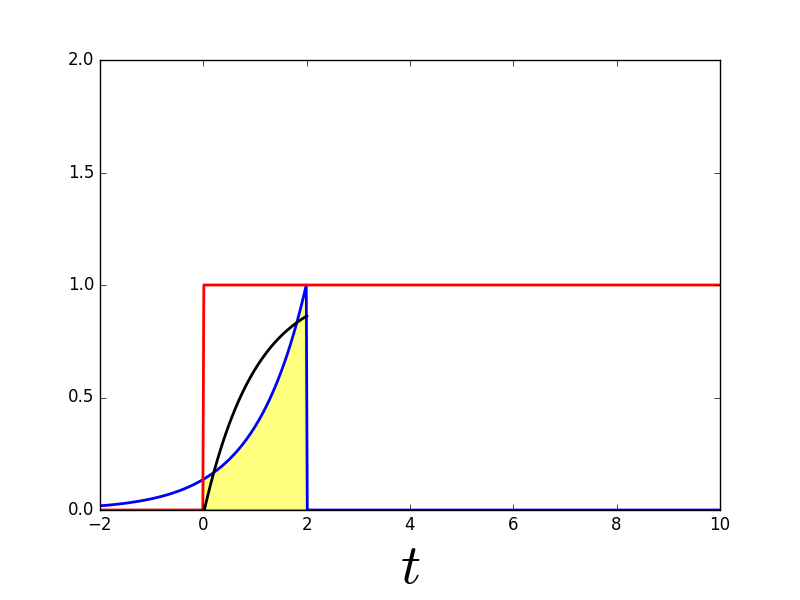}
\caption{$t = 2$.}
\end{subfigure}
\begin{subfigure}[t]{0.32\textwidth}
\includegraphics[width=1.0\linewidth]{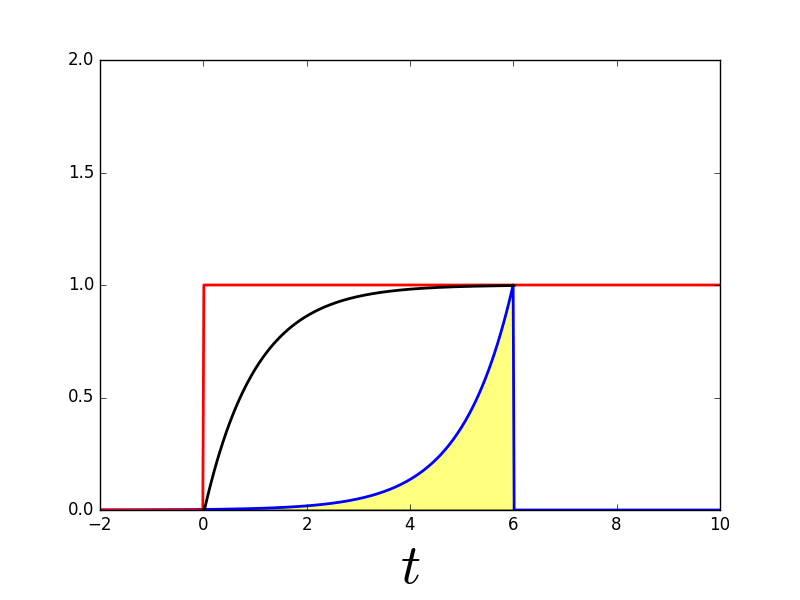}
\caption{$t = 6$.}
\end{subfigure}
\end{center}
\caption{ Evolution of the convolution integral  $\int_0^t f(s) e^{-(t-s)} ds$. To evaluate the convolution integral graphically, first reflect $e^{t}$, add a time offset, and then slide it along the t-axis. Then $f(t)$ is plotted as a function of $t$. The integral is the area under the curve of $e^{(t-s)}f(t)$.}
\label{fig:LinearConvolved}
\end{figure}

Significant insight is gained from Eq.~\ref{eq:linearOrthoSolution}. It is seen that the convolution memory integrand contains the exponential term $e^{\Lambda s}$. When the eigenvalues are negative (a general characteristic of stable systems), the exponential operator leads to the integrand having finite support. The timescale of this support is proportional to the inverse of the eigenvalues. To explain this, a  pictorial representation of the evaluation of a simple convolution integral is given in Figure~\ref{fig:LinearConvolved}. The figures show the graphical evaluation of the convolution integral $\int_0^t f(s) e^{\lambda(t-s)} ds$ with $f(t) = H(t)$ and $\lambda = -1$. 
It is seen that the exponential operator limits the support of the integrand. The time scale of this support is related to the argument of the exponential operator, which is related to the eigenvalues of the auxiliary system.

\subsection{Modeling of the Memory Kernel}\label{sec:modelMemory}
To obtain a reduction in complexity, a surrogate to the memory must be devised.
The t-model is perhaps the simplest approximation to the memory term and can be obtained by expanding the memory integrand in a Taylor series about $s = 0$ and retaining only the leading term,
\begin{equation}\label{eq:tmodel}
 \MC{P}  \int_0^t e^{{(t - s)}\mathcal{L}} \mathcal{P}\mathcal{L} e^{s \mathcal{Q} \mathcal{L}} \MC{QL}\phi_{0j} ds \approx 
  \MC{P}\int_0^t e^{t \mathcal{L}} \MC{PLQL} \phi_{0j} ds =  t \MC{PLQL}\phi_j(t).
\end{equation}
Various interpretations of the t-model exist, the most common of which is that the memory is infinitely long. The t-model can be expected to be accurate when minimal flow is invoked by the orthogonal dynamics, i.e. $e^{s \MC{QL}}\MC{QL}\phi_{0j} \approx \MC{QL}\phi_{0j}$. Its accuracy additionally requires that the flow invoked by the true dynamics $e^{t \MC{L}}$ is slow to evolve in time. Despite its simplicity, the t-model has been successfully applied to Burgers equation and as a sub-grid model in Large Eddy Simulations~\cite{ChandyFrankelLES}.

The insight gained from Section~\ref{sec:ortho} suggests that the memory integrand has a finite support. Based on this, a more general class of models that assume a finite support of the memory integrand is now considered. The models derived here were first considered by  Stinis~\cite{stinisEuler,stinis_finitememory}. For notational purposes, define 
\begin{equation}\label{eq:M-Z_FM1}
w^{(m)}_{j}(\phi_0,t) = 
 \MC{P} \int_{a_m(t)}^t e^{{s}\mathcal{L}} \mathcal{P}\mathcal{L} e^{(t-s) \mathcal{Q} \mathcal{L}} \big(\MC{QL} \big)^{m+1}\phi_{0j} ds,
\end{equation}
where $a_m(t) = t - \tau_m(t)$. Note the change of variables $t' = t-s$. For clarity of presentation, the dependence of $\hat{\phi}$ on the initial conditions $\phi_0$ and time $t$ will be implicitly assumed throughout the rest of this section. Setting $m=0$ (in which case Eq.~\ref{eq:M-Z_FM1} is simply the memory term) and differentiating Eq.~\ref{eq:M-Z_FM1} with respect to time yields
\begin{equation}\label{eq:M-Z_FM1_2}
\frac{d}{dt}  w^{(0)}_{j}(\phi_0,t) = 
 e^{t \MC{L}}\MC{PLQL}{\phi}_{0j} -  e^{(t-\tau_0)\MC{L}}\MC{PL}e^{\tau_0\MC{QL}}\MC{QL}{\phi}_{0j} a_0'(t)
  + \MC{P} \int_{a_0(t)}^t  e^{{s}\mathcal{L}} \mathcal{P}\mathcal{L} e^{(t-s) \mathcal{Q} \mathcal{L}} \MC{QLQL}\phi_{0j} ds .
\end{equation}
Note that the first term on the right hand side does not require the solution of the orthogonal dynamics equation. The second term on the right hand side is dependent on the orthogonal dynamics.  This dependence can be eliminated by using a discrete integration scheme to express the memory integral. In the case of the trapezoidal rule
$$w^{(0)}_{j}(t) = \bigg[  e^{t \MC{L}}\MC{PLQL}{\phi}_{0j} + 
e^{(t-\tau_0)\MC{L}}\MC{PL}e^{\tau_0\MC{QL}}\MC{QL}{\phi}_{0j}  \bigg] 
 \frac{\tau_0(t)}{2} + \MC{O}(\tau_0^2).$$
In order to handle the case where the memory length $\tau_0$ is not necessarily small, the memory integral in Eq.~\ref{eq:M-Z_FM1} is partitioned into $N$ sub-intervals,
\begin{multline}
\MC{P} \int_{t - \tau_0}^t e^{{s}\mathcal{L}} \mathcal{P}\mathcal{L} e^{(t-s) \mathcal{Q} \mathcal{L}} \MC{QL}\phi_{0j} ds = \MC{P} \int_{t - \Delta \tau_0}^t e^{{s}\mathcal{L}} \mathcal{P}\mathcal{L} e^{(t-s) \mathcal{Q} \mathcal{L}} \MC{QL}\phi_{0j} ds + \\ 
\MC{P} \int_{t - 2 \Delta \tau_0}^{t -  \Delta \tau_0} e^{{s}\mathcal{L}} \mathcal{P}\mathcal{L} e^{(t-s) \mathcal{Q} \mathcal{L}} \MC{QL}\phi_{0j} ds +  \ldots\ + 
\MC{P} \int_{t - N \Delta \tau_0}^{t - (N-1) \Delta \tau_0} e^{{s}\mathcal{L}} \mathcal{P}\mathcal{L} e^{(t-s) \mathcal{Q} \mathcal{L}} \MC{QL}\phi_{0j} ds, 
\end{multline}
where $\Delta \tau_0 = \tau_0/N$. 
Define
$$w_j^{(m,n)} = \MC{P} \int_{t - n \Delta \tau_0}^{t - (n-1) \Delta \tau_0} e^{{s}\mathcal{L}} \mathcal{P}\mathcal{L} e^{(t-s) \mathcal{Q} \mathcal{L}} \big( \MC{QL} \big)^{m+1} \phi_{0j} ds$$
for $n=1,2,...,N$. 
Applying the trapezoidal integration scheme to each sub-interval yields the general form
\begin{multline}\label{eq:M-Z_FM1_3}
\frac{d}{dt} w_{j}^{(0,n)}(\phi_0,t) =  \bigg[\sum_{i=1}^{n-1} (-1)^{n+i+1} w_{j}^{(0,i)} \bigg] \frac{2}{\Delta \tau_0} \bigg(2 - (2n-1) \Delta \tau_0' \bigg) - \frac{2}{\Delta \tau_0} w_{j}^{(0,n)} \big(1 - n \Delta \tau_0' \big) + \\
\big(2 - (2n-1) \Delta \tau_0' \big)e^{t \MC{L}}\MC{PLQL}{\phi}_{0j} 
   + \MC{P} \int_{t - n \Delta \tau_0}^{t - (n-1) \Delta \tau_0}  e^{{s}\mathcal{L}} \mathcal{P}\mathcal{L} e^{(t-s) \mathcal{Q} \mathcal{L}} \MC{QLQL}\phi_{0j} ds + 
    \MC{O}(\Delta \tau_0^2).
\end{multline}
The right hand side of Eq.~\ref{eq:M-Z_FM1_3} is now closed with the exception of the memory term. Assume that the new memory term has a finite support from $t - \tau_1(t)$ to $t$. A differential equation for $w^{(1,n)}$ can be developed by again differentiating the convolution integral in Eq.~\ref{eq:M-Z_FM1_3} with respect to time and using the trapezoidal rule. 
The differentiation process can be continued to build an infinite hierarchy of Markovian equations. The general form obtained is
\begin{multline}\label{eq:M-Z_FM1_6}
\frac{d}{dt} w_{j}^{(m,n)}(\phi_0,t) =  \bigg[\sum_{i=1}^{n-1} (-1)^{n+i+1} w_{j}^{(m,i)} \bigg] \frac{2}{\Delta \tau_m} \bigg(2 - (2n-1) \Delta \tau_m' \bigg) - \frac{2}{\Delta \tau_m} w_{j}^{(m,n)} \big(1 - n \Delta \tau_m' \big) + \\
\big(2 - (2n-1) \Delta \tau_m' \big)e^{t \MC{L}}\MC{PL}\big(\MC{QL}\big)^{m+1}{\phi}_{0j} 
 + \MC{P} \int_{t - n \Delta \tau_m}^{t - (n-1) \Delta \tau_m}  e^{{s}\mathcal{L}} \mathcal{P}\mathcal{L} e^{(t-s) \mathcal{Q} \mathcal{L}} (\MC{QL})^{m+2}\phi_{0j} ds + 
    \MC{O}(\Delta \tau_m^2).
\end{multline}
The infinite hierarchy of equations must be truncated at some point. This can be done by modeling the effects of $w_j^{(m+1,n)}$ or, more simply, by neglecting it. Neglecting $w_j^{(m+1,n)}$ can be justified if the support or magnitude of the integrand decreases with the repeated application of $\MC{QL}$. The derivation above can be carried out using higher order quadrature~\cite{stinis_finitememory}. In this work, models with a constant memory length and one sub-interval are considered. In this case the models simplify to
\begin{equation}\label{eq:M-Z_FM1_7}
\frac{d}{dt} w_{j}^{(m)}(\phi_0,t) =  -\frac{2}{\tau_m}w^{(m)}_{j}(t) + 2 e^{t \MC{L}} \MC{PL}\big(\MC{QL}\big)^{m+1}\phi_{0j} + 
w_j^{(m+1)}.
\end{equation}

\section{Application to Burgers Equation}
The viscous Burgers equation (VBE) is a one-dimensional equation that serves as a toy-model for turbulence. The VBE has been well-studied and is a canonical problem to test the performance of sub-grid models. It should be noted that solutions of the VBE are not chaotic, a property that is one of the defining features of turbulence. 
The VBE in Fourier space is given by
\begin{equation}\label{eq:VBE_freq}
\frac{\partial u_k}{\partial t} + \frac{\imath  k}{2} \sum_{\substack{ p + q = k \\ p ,q \in F \cup G }}u_p u_{q} = -\nu k^2 u_k, \qquad k \in F \cup G
\end{equation}
with $u_k(0) = u_{0k}$. The Fourier modes $u = \{\hat{u},\tilde{u} \}$ are contained within the union of two sets, $F$ and $G$. In the construction of the reduced order model, the resolved modes are $\hat{u} \in F$ and the unresolved modes are $\tilde{u} \in G$. Partitioning Eq.~\ref{eq:VBE_freq} into the resolved and unresolved sets, the evolution equation for the resolved variables is written as
\begin{equation}\label{eq:LESVBE_freq}
\frac{\partial {u}_k}{\partial t} + \frac{\imath k}{2} \sum_{\substack{ p + q = k \\ p \in F ,q \in F }}{u}_{p}{u}_{q} = -\nu k^2 {u}_k -  \frac{\imath k}{2}\bigg( \sum_{\substack{ p + q = k \\ p \in G ,q \in G }}{u}_{p} {u}_{q} +   \sum_{\substack{ p + q = k \\ p \in F ,q \in G }}{u}_{p} {u}_{q}  +\sum_{\substack{ p + q = k \\ p \in G ,q \in F }}{u}_{p} {u}_{q}  \bigg) \qquad k \in F . 
\end{equation}
Eq.~\ref{eq:LESVBE_freq} is equivalent to the LES form of the VBE with a sharp spectral cutoff filter. Note that the sub-grid stress in Fourier space is written as 
$${\tau}_k^{SGS} = \frac{1}{2} \bigg( \sum_{\substack{ p + q = k \\ p \in G ,q \in G }}{u}_{p} {u}_{q} +   \sum_{\substack{ p + q = k \\ p \in F ,q \in G }}{u}_{p} {u}_{q}  +  \sum_{\substack{ p + q = k \\ p \in G ,q \in F }}{u}_{p} {u}_{q}  \bigg) .$$
 The last term on the RHS of Eq.~\ref{eq:LESVBE_freq} represents the effect of the unresolved scales on the resolved scales and must be modeled. 

Traditional sub-grid models make an isotropic eddy viscosity approximation, where the sub-grid stress is assumed to be linearly proportional to the strain-rate of the resolved scales. In such models the sub-grid stress is modeled in the physical domain by
$$\tau_{ij}^{SGS} = -2 \nu_{sgs} {S}_{ij},$$
with $S$ being the resolved strain-rate tensor. The eddy viscosity $\nu_{sgs}$ is determined from the filtered flow field. The Smagorinsky~\cite{Smagorinsky} model is perhaps the most notable sub-grid model and uses
$$\nu_{sgs} = (C_s \Delta )^2 |\tilde{S}|.$$
The static and dynamic Smagorinsky models will be used as a reference  to compare the Mori-Zwanzig based closures.

\subsection{Construction of Mori-Zwanzig Models} 
Mori-Zwanzig closures based on the expectation projection are considered.
Let $f \in L^2$. The projection of $f$ onto $\hat{L}^2$ is given by 
\begin{equation}\label{eq:conditionalExpectation}
(\MC{P}f)(\hat{\phi}_0) = E[f | \hat{\phi}_0] = \frac{\int f(\phi_0)\rho(\phi_0) d\tilde{\phi}_0}{\int \rho(\phi_0) d \tilde{\phi}_0}.
\end{equation} 
 The density of the initial conditions is assumed to consist of independent Gaussian distributions in the zero-variance limit as in Ref.~\citenum{bernsteinBurgers}. For this density, the expectation projection sets all unresolved modes to be zero, {\em i.e.}
$$\MC{P}\big(f(\hat{\phi}_0, \tilde{\phi}_0)\big) = f(\hat{\phi}_0,0).$$
The high fidelity model is taken to have support $-N \le k \le N-1$, while the reduced order model has support for $-N/2 \le k \le N/2-1$. Evaluating Eq.~\ref{eq:M-Z_Identity3} for the VBE and casually commuting the non-linear Markovian term yields 
\begin{equation}\label{eq:M-ZVBE_freq}
\frac{\partial {u}_k}{\partial t} + \frac{\imath k}{2} \sum_{\substack{ p + q = k \\ p \in F ,q \in F }}{u}_{p} {u}_{q} = -\nu k^2 {u}_k + \MC{P}\int_0^t K( \mathbf{{u}}(t),t-s)  ds.
\end{equation}
Evaluation of the convolution integral in Eq.~\ref{eq:M-ZVBE_freq} is not tractable, and it is estimated with the models previously discussed. For the VBE, finite memory models for $m=1,2,3$ are considered; as well as the t-model. One sub-interval is used in the trapezoidal approximation. For the VBE the first order term is found to be
\begin{equation}\label{eq:FM1_gen}
 e^{t \MC{L}} \MC{PLQL}u_{0k}= 
 -\imath k \sum_{\substack{ p + q = k \\ p \in F ,q \in G }}u_{p}  \bigg[-\frac{\imath q}{2}\sum_{\substack{ r + s = q \\ r,s \in F  }}u_{r} u_{s} \bigg] 
 \qquad k \in F.
 \end{equation}
The form of the required terms in the higher order models are given in the Appendix. 
Throughout the remainder of this manuscript, the finite memory model based on the first order expansion (Eq.~\ref{eq:FM1_gen}) will be referred to as the first order finite memory model, FM1. Similarly, the second and third order expansions will be referred to as FM2 and FM3. 

\subsection{Numerical Implementation}
The VBE is solved numerically using a Fourier-Galerkin spectral method. 
The FFT calculations are padded by the 3/2 rule. An explicit low storage 4th order Runge-Kutta method is used for time integration. 

Numerical simulations of the VBE are performed with the initial  condition~\cite{ZJWangBurgers}
\begin{equation}\label{eq:burgersIC}
u(x) = U_0^* \sum_{i=1}^{k_c} \sqrt{2 E(k_i)} \sin(k_i x + \beta_i),
\end{equation}
where $E(k) = 5^{-5/3}$ if $1\le k \le5$ and $E(k) = k^{-5/3}$ for $k > 5$. Eq.~\ref{eq:burgersIC} initializes an energy spectrum with a -5/3 slope for $k > 5$. The phase angle $\beta$ is a random number in the domain $[-\pi,\pi]$. A constant seed value is used in all of the simulations. No energy is added to the flow after a cut-off frequency $k_c$, such that LES simulations are initially fully resolved.

\subsubsection{Selection of the Memory Length}\label{sec:selectMemoryLength}
\begin{figure}
 \begin{centering}
  \captionsetup{justification=justified,singlelinecheck=false}
  \includegraphics[trim={0.05cm 0 1.75cm 1cm},clip,width=.55\linewidth]  {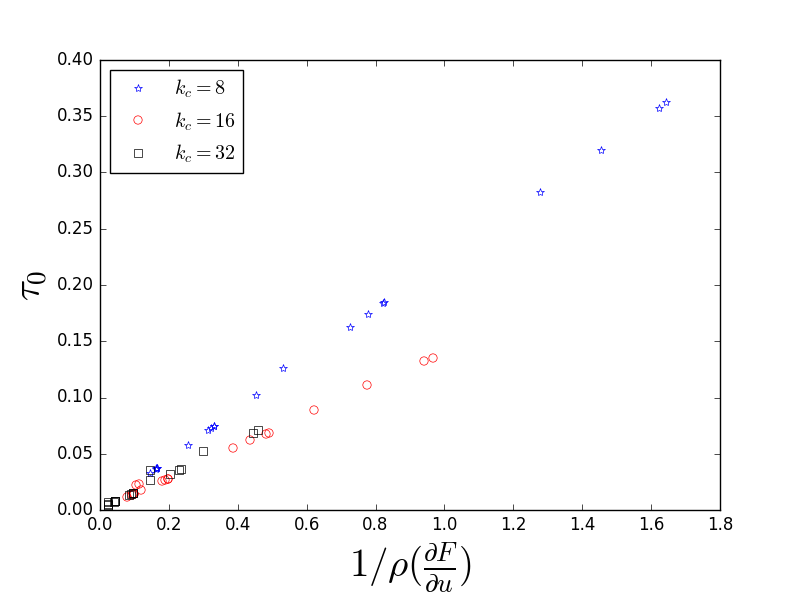}
  \caption{MAP solution for the first order model time constant $\tau_0$ plotted against the inverse of the spectral radius of the Jacobian of the resolved variables.}
  \label{fig:eigs} 
\end{centering}
\end{figure}

To determine the memory length, one would ideally like to directly compute the memory kernel. Computing the memory kernel involves solving the orthogonal dynamics equation and is not directly tractable. In this work, an intuitive approach is considered. In Section~\ref{sec:ortho} it was demonstrated that, in the linear case, the memory length could be related to the eigenvalues of the auxiliary dynamical system that solves the orthogonal dynamics. Since Burgers equation does not exhibit scale separation, a logical hypothesis is that a mean time scale can be related to the spectral radius of the Jacobian of the resolved variables
$$ \tau \propto 1/ \rho \bigg( \frac{\partial \mathbf{R}}{\partial \mathbf{u}}  \bigg).$$
To provide evidence for this argument, a parametric study was performed involving 60 cases. The simulations were initialized with Eq.~\ref{eq:burgersIC} and operated over a range of Reynolds numbers and resolutions. The cases considered were permutations of the following parameters: $\nu=[0.05,0.01,0.005,0.001,0.0005], k_c = [8,16,32], U_0^* = [1,2,5,10]$. The DNS simulations were carried out using 4096 resolved modes. For each case, the time constant $\tau_0$ in the first order model is found by solving an inverse problem. For simplicity, $\tau_0$ is taken to be constant in time. The optimal time constant in the least squares sense was found by minimizing the difference of the total kinetic energy dissipation rate between the reduced order model solution and a high resolution DNS solution. The solution was minimized for $t \in [0,2]$ using data at discrete time-steps spaced by intervals of $\Delta t = 0.01.$ The discrete penalty function is given by
$$\mathcal{J} = \sum_{n=1}^N \bigg( \left[\frac{dK_n}{dt}\right]_{M-Z} -  \left[ \frac{dK_n}{dt} \right]_{DNS} \bigg)^2$$
where $N = 2/0.01 = 2000$. The penalty function was minimized using SciPy's optimization suite. A downhill simplex algorithm was used. It is noted that the inferred results were similar to a penalty function that minimized the difference in total kinetic energy. Figure~\ref{fig:eigs} shows the inferred time constants plotted against the temporal mean of the spectral radius of the Jacobian of the resolved variables. The $k_c = 16$ and $k_c = 32$ cases are seen to collapse to the same line. The $k_c=8$ cases also collapse to a linear line, but with a slightly greater slope. Given the wide range of cases considered, the collapse of the data is quite good. This result suggests that the properties of the Jacobian of the resolved variables can be used as a good indicator of the memory length. A fit of the above data yields $\tau_0 \approx 0.2 / {\rho \big(\frac{\partial \mathbf{R}}{\partial \mathbf{u}_0} \big) }.$ A more rigorous fitting process and the effect of uncertainty in the time constant is provided in Appendix~\ref{sec:appendix1}.

Before proceeding further, we make several comments. First, the results of this section should be viewed as evidence rather than proof that a finite memory approximation is appropriate for the Burgers equation. The first order model has errors that are due to the trapezoidal integration as well as neglecting $w_j^{(1)}$. 
We note that the inferred memory length should not be viewed as a physical parameter. Nonetheless, the memory length \textit{has a physical meaning} and the inferred results are consistent with that assertion. An increased Reynolds number leads to a decreased time scale, while a decrease in numerical resolution leads to an increase in time scale. This makes the time constant different than a heuristic tuning constant.

\subsection{Numerical Results} 
\begin{figure}
\begin{subfigure}[t]{1.\textwidth}
  \captionsetup{justification=justified,singlelinecheck=false}
  \includegraphics[trim={0.05cm 0 1.75cm 1cm},clip,width=.5\linewidth]  {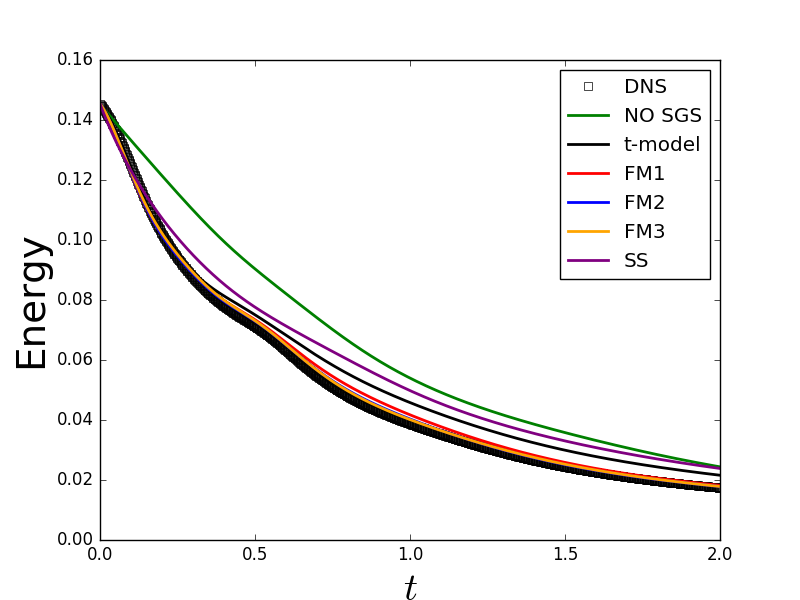}
  \includegraphics[trim={0.05cm 0 1.75cm 1cm},clip,width=.5\linewidth]  {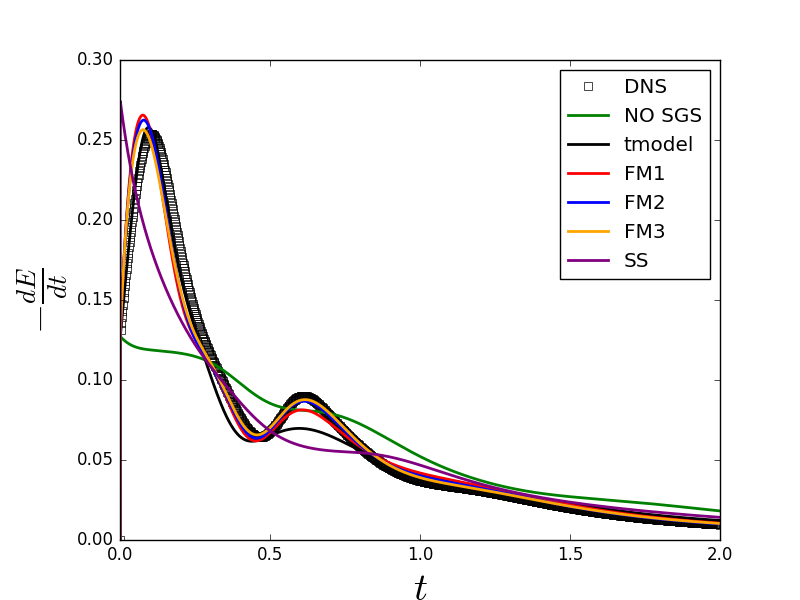}
  \caption{ Temporal evolution of total kinetic energy (left) and rate of decay of kinetic energy (right).}
  \label{fig:Burgers_LES1a} 
\end{subfigure}

\begin{subfigure}[t]{1.\textwidth}
  \captionsetup{justification=justified,singlelinecheck=false}
  \includegraphics[trim={0.05cm 0 1.75cm 1cm},clip,width=.5\linewidth]  {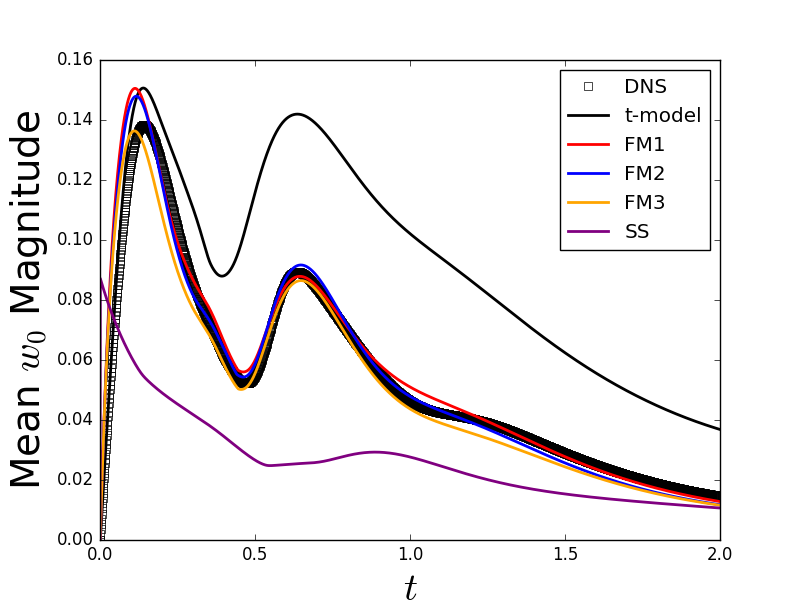}
  \includegraphics[trim={0.05cm 0 1.75cm 1cm},clip,width=.5\linewidth]  {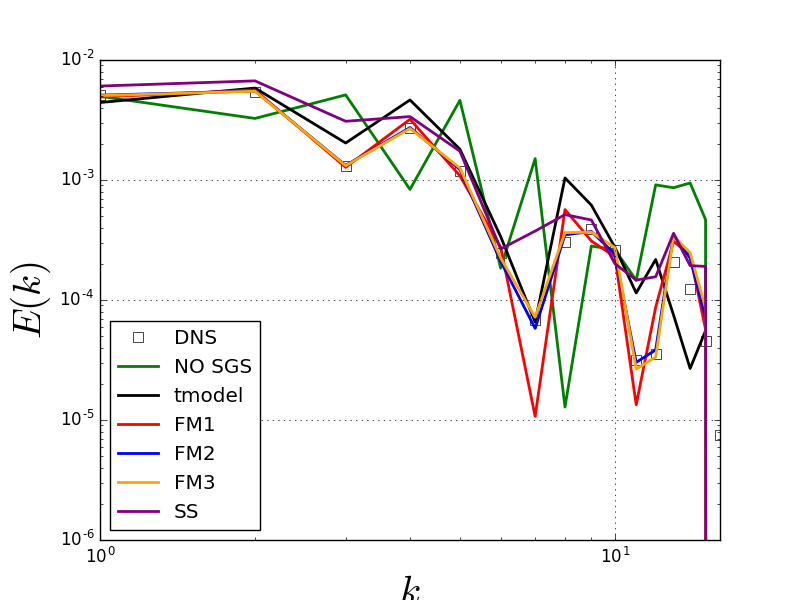}
\caption{ Temporal evolution of the mean magnitude of ${w^{(0)}}$ (left) and energy spectrum at ${t = 2.0}$. Note that $w^{(0)} = {ik \tau_{sgs}}$.}
\label{fig:Burgers_LES1b} 
\end{subfigure}
\caption{ A comparison of Large Eddy Simulations performed with 32 resolved modes to filtered DNS data obtained from a simulation performed with 2048 resolved modes.}
\label{fig:Burgers_LES1} 
\end{figure}

\begin{figure}
\begin{subfigure}[t]{1.\textwidth}
  \captionsetup{justification=justified,singlelinecheck=false}
  \includegraphics[trim={0.05cm 0 1.5cm 1cm},clip,width=.5\linewidth]  {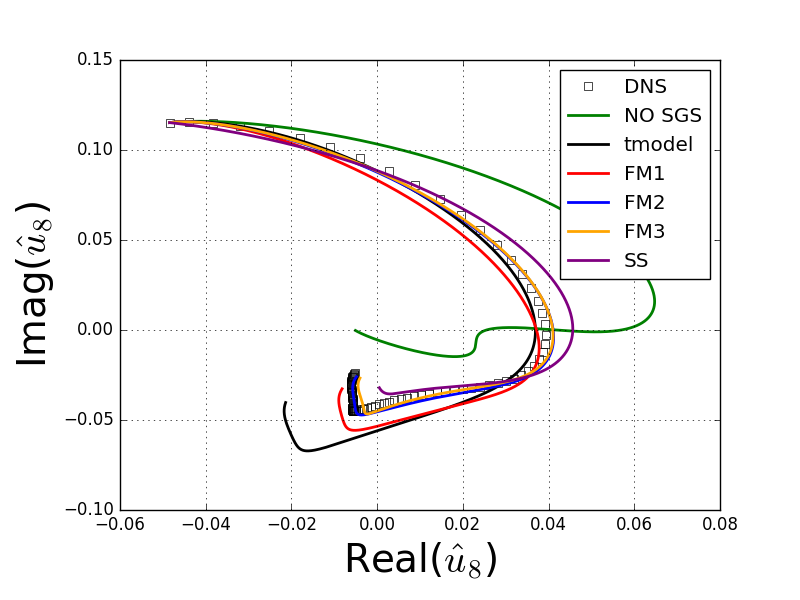}
  \includegraphics[trim={0.05cm 0 1.5cm 1cm},clip,width=.5\linewidth]  {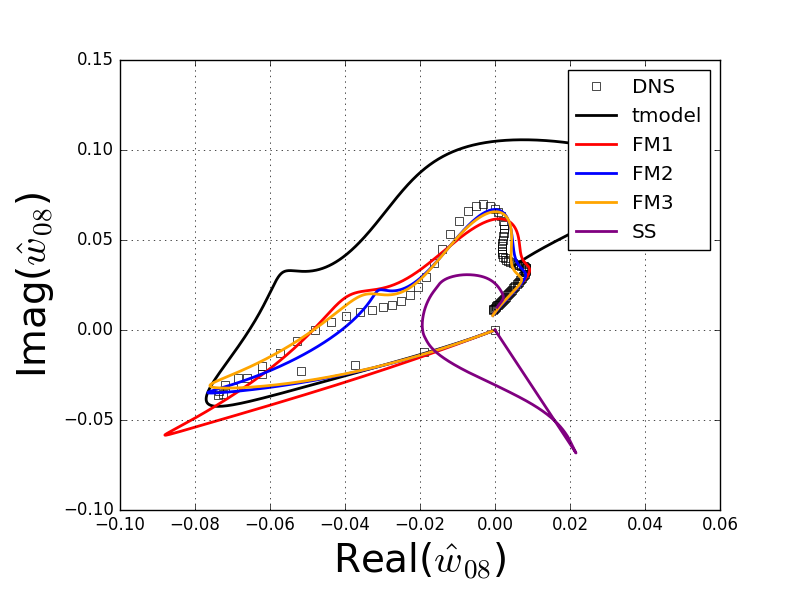}
  \caption{ Evolution of the eighth mode of ${u}$ (left) and ${w}$ (right).}
  \label{fig:Burgers_LES2a} 
\end{subfigure}

\begin{subfigure}[t]{1.\textwidth}
  \captionsetup{justification=justified,singlelinecheck=false}
  \includegraphics[trim={0.05cm 0 1.5cm 1cm},clip,width=.5\linewidth]  {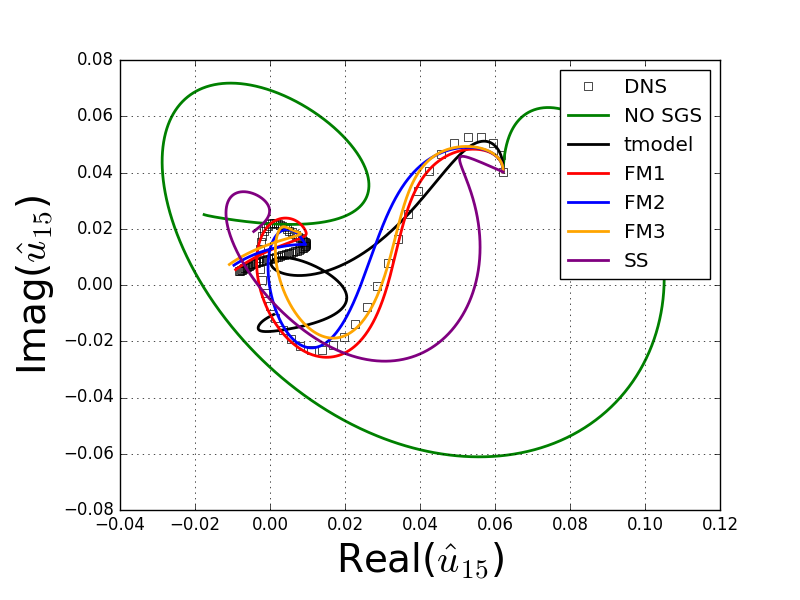}
  \includegraphics[trim={0.05cm 0 1.5cm 1cm},clip,width=.5\linewidth]  {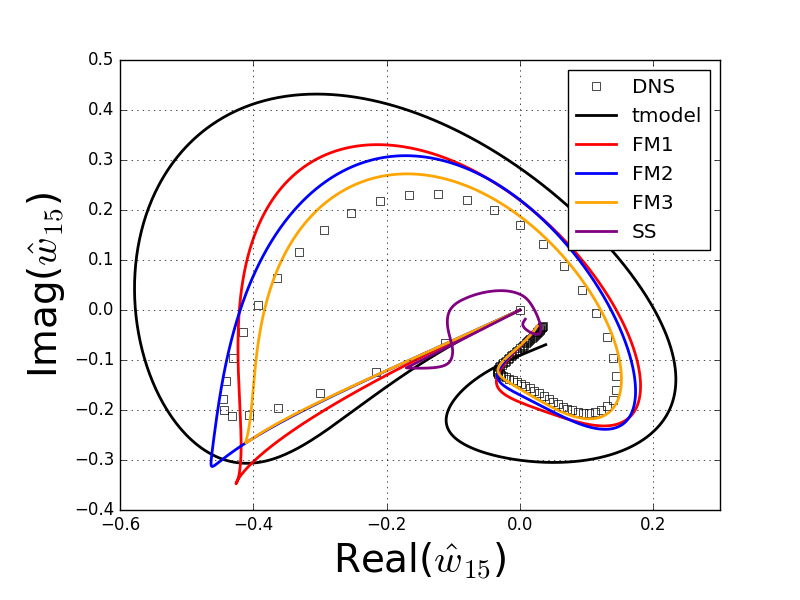}
  \caption{ Evolution of the 15th mode of ${u}$ (left) and ${w}$ (right).}
  \label{fig:Burgers_LES2b} 
\end{subfigure}
\caption{  Evolution of select modes of ${u}$ and ${w}$ in phase space. In phase space the DNS data (denoted by $\square$) is sparse around $t=0$ and becomes clustered as $t \rightarrow 2$. }
\label{fig:Burgers_LES2} 
\end{figure}

 Direct numerical simulations of the VBE were performed with $2048$ resolved modes ($-1024\le k \le 1023$) on a spatially periodic domain of length $2\pi$ for $t \in [0,2]$. The initial condition given in Eq.~\ref{eq:burgersIC} is used with $U_0^* = 1$ and $\nu = 0.01$.
LES  is performed with the SGS models described above. The LES simulations are initialized with the DNS solution for $k \le k_c$ such that the initial condition of the LES  is fully resolved. The simulations were performed with 32 resolved modes, corresponding to a cut-off frequency at $k_c = 16$. The memory length $\tau_0$ was selected by the procedure described in the previous section. A formal estimation procedure was not used for the memory lengths of the higher order models, which were simply chosen to be $\tau_{1} = \tau_2 = 0.5 \tau_0$. A summary of the relevant computational details is given in Table~\ref{table:Burgers1_summary}.
\begin{table}
\begin{center}
    \begin{tabular}{| l | l | l | l | l | l | l |}
    \hline
     & DNS & Smagorinsky & t-model & FM1 & FM2 & FM3  \\ \hline
    $N$ & $2048$ & $32$ & $32$ & $32$ & $32$ & $32$\\ \hline
$\Delta t$& 1e-4 & 1e-3 & 1e-3 & 1e-3 & 1e-3 & 1e-3\\ \hline
Constants & NA & $Cs=0.2$ & $\tau_0=t$ & $\tau_0 = 0.135$ & $\tau_{0,1}=0.135,0.07$ & $\tau_{0,1,2}=0.135,0.07,0.07$ \\ \hline
    \end{tabular}
\end{center}
\caption{Summary of computational details for the numerical experiments of Burgers equation.}
\label{table:Burgers1_summary}
\end{table}
Figures~\ref{fig:Burgers_LES1} and~\ref{fig:Burgers_LES2} compare the Mori-Zwanzig based models to the filtered DNS data, the Smagorinsky model, and a simulation ran on a 32 point mesh without any sub-grid model. Figure~\ref{fig:Burgers_LES1a} shows the temporal evolution of the total kinetic energy and rate of kinetic energy decay. Figure~\ref{fig:Burgers_LES1b} shows the temporal evolution of the mean magnitude of $w^{(0)}$ (note $w^{(0)} = i k \tau^{sgs}$) and the energy spectrum at $t = 2.0$. Figure~\ref{fig:Burgers_LES2} shows the trajectories of the 8th (Fig.~\ref{fig:Burgers_LES2a}) and 15th (Fig.~\ref{fig:Burgers_LES2b}) modes of $u$ and $w^{(0)}$ in the complex plane. A brief discussion of the results of each simulation will now be provided.

The simulation performed without an SGS model severely under-predicts the rate of energy decay at early time, leading to an over-prediction in total kinetic energy. As expected, the simulation under-predicts the dissipation rate. 
With no sub-grid mechanism present to remove energy from high wave numbers, a pile-up of energy is seen for high $k$, as evidenced in Figure~\ref{fig:Burgers_LES1b}. This phenomena indicates that the simulation is under-resolved and a sub-grid model is indeed required. The evolution of the individual modes of $\hat{u}$ contain significant error, particularly around the cutoff frequency. The evolution of the 15th mode, as shown in Figure~\ref{fig:Burgers_LES2a}, is an excellent example of the error present in high wave numbers.

The Smagorinsky model offers  improvements. The simulation utilizing the basic SGS model provides decent predictions for both the total kinetic energy and the dissipation of kinetic energy. The energy spectrum at $t = 2.0$ and trajectories of the individual modes are additionally much improved.  However, the Smagorinsky model is unable to differentiate the resolved scales from the unresolved scales. Despite being completely resolved at $t=0$, the Smagorinsky model predicts that the sub-grid content is maximum at $t=0$. The resulting predictions for $w^{(0)}$ are both quantitatively and qualitatively incorrect. In particular, the individual trajectories of ${w}^{(0)}$ show no similarity to that of the DNS. It is recognized that the Smagorinsky model was developed for homogeneous turbulent flows in the physical domain, so a critical evaluation of the model on the Burgers equation is not completely appropriate.
 
Simulations performed using the t-model provide improved predictions. The largest error is present in the prediction for $w^{(0)}$, where it is seen that the t-model slightly over predicts sub-grid content (especially for $t > 0.5$), but the predictions are still qualitatively correct. The model correctly predicts the initial peak in $dE/dt$ and $w^{(0)}$ around $t = 0.15$ and qualitatively shows the presence of the second peak around $t = 0.6$. The trajectories of the individual modes in ${u}$ and ${w}^{(0)}$ are improved, but become less accurate for late time. The prediction for the energy spectrum at $t=2.0$ is not noticably better than that predicted by the Smagorinsky model. The explicit presence of $t$ in the model leads to substantial error for large time. The performance of the t-model for the VBE shows the merit in the M-Z-based models. To reiterate, the t-model contains no heuristics or  coefficients. Work by Stinis ~\cite{RenormalizedMZ} and our own numerical experiments show that re-normalization of the t-model can lead to more accurate results. 

The finite memory models provide relatively accurate predictions for all quantities. The evolution of total kinetic energy, dissipation of kinetic energy, and mean sub-grid predictions are in good agreement with the DNS. The first order finite memory model accurately predicts the instantaneous energy spectrum at $t = 2.0$ for low wave numbers, while the second and third order models provide accurate predictions for all wave numbers. The trajectories of the individual modes are close to that of the DNS, as are the trajectories for the sub-grid-terms. 

\begin{figure}

  \captionsetup{justification=justified,singlelinecheck=false}
  \includegraphics[trim={0.05cm 0 1.5cm 1cm},clip,width=.49\linewidth]  {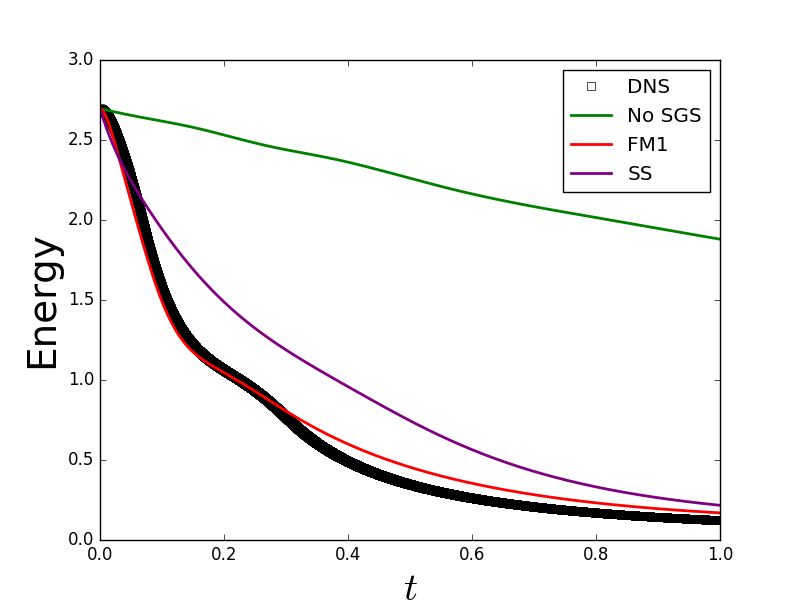}
  \includegraphics[trim={0.05cm 0 1.5cm 1cm},clip,width=.49\linewidth]  {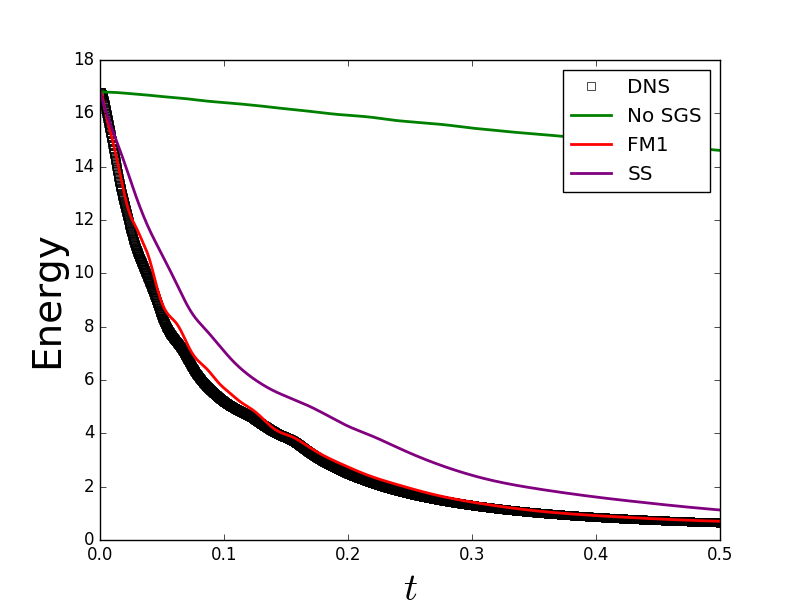}
  \includegraphics[trim={0.05cm 0 1.5cm 1cm},clip,width=.49\linewidth]  {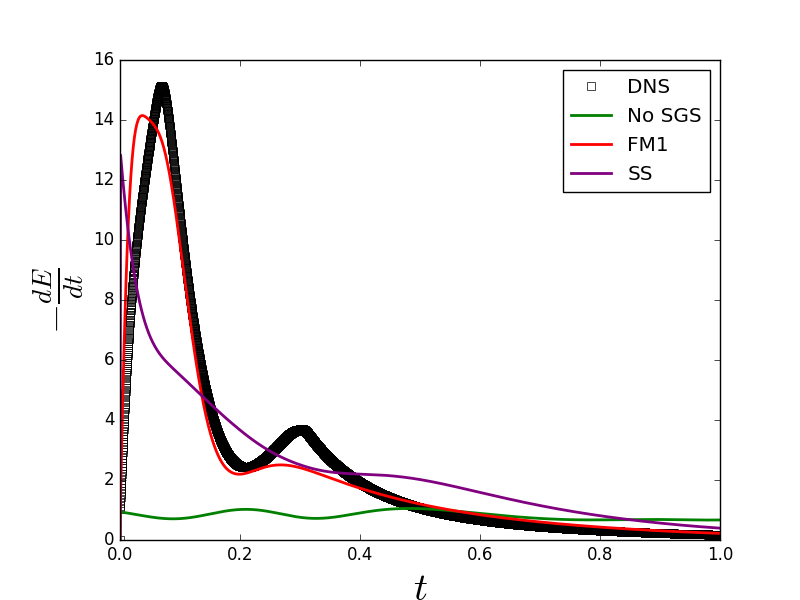}
  \includegraphics[trim={0.05cm 0 1.5cm 1cm},clip,width=.49\linewidth]  {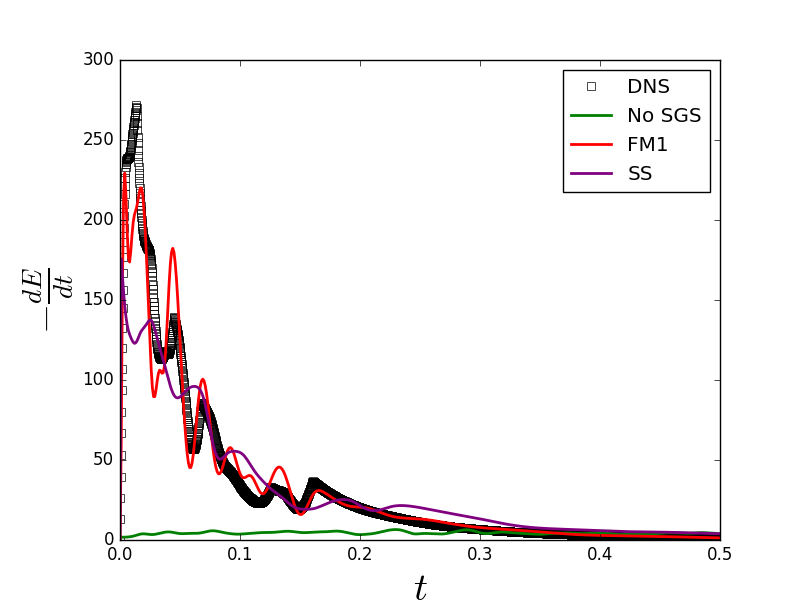}
  \includegraphics[trim={0.05cm 0 1.5cm 1cm},clip,width=.49\linewidth]  {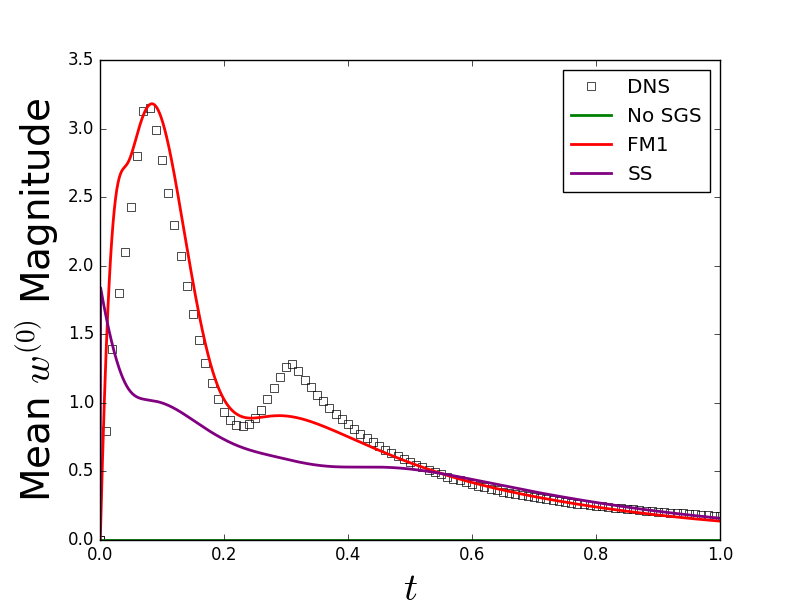}
  \includegraphics[trim={0.05cm 0 1.5cm 1cm},clip,width=.49\linewidth]  {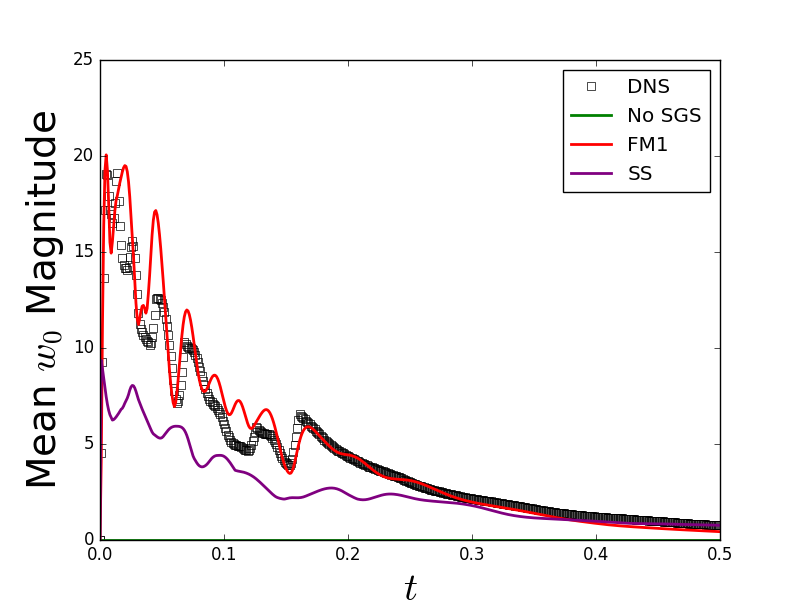}
\caption{ Evolution of select quantities of the additional simulations of the VBE. The conditions are $k_c=8, \nu=0.01, U_0^*=5$ (left) and $k_c=32, \nu=5 \times 10^{-4}, U_0^*=10$. }
\label{fig:Burgers_LES3} 
\end{figure}

Results of the first order finite memory model are shown in Figure~\ref{fig:Burgers_LES3} for two additional cases. The first is run at a resolution of $k_c=8$, a viscosity of $\nu = 0.01$, and a scaling of $U_0^* = 5$. The second case is run at a resolution of $k_c=32$, a viscosity of $\nu = 5 \times 10^{-4}$, and a scaling of $U_0^*=10$. The time constants were again selected by the scaling of the spectral radius of the Jacobian. Both of these cases are significantly under-resolved.  The low viscosity case in particular has numerous shocks and required 4096 resolved modes for the DNS calculation.

\section{Application to the Triply Periodic Navier-Stokes Equations}
Coarse-grained simulations of the Navier-Stokes equations are now considered. The incompressible Navier-Stokes equations in Fourier space are given by
\begin{equation}\label{eq:NSspec}
\bigg(\frac{\partial}{\partial t} + \nu k^2 \bigg){u}_{i} (\mathbf{k},t) + \bigg(\delta_{im} - \frac{k_i k_m}{k^2} \bigg) \imath k_j \sum_{\substack{ \mathbf{p} + \mathbf{q} = \mathbf{k} \\ \mathbf{p,q} \in F \cup G } } {u}_j (\mathbf{p},t)  {u}_m (\mathbf{q},t) = 0  \qquad \mathbf{k} \in F \cup G,
\end{equation}
where the Fourier modes  again belong to the union of two sets. Separating the modes into the resolved and unresolved sets yields,
\begin{equation}\label{eq:NSspec_filtered}
\bigg(\frac{\partial}{\partial t} + \nu k^2 \bigg){u}_{i} (\mathbf{k},t) + \bigg(\delta_{im} - \frac{k_i k_m}{k^2} \bigg) \imath k_j \sum_{\substack{ \mathbf{p} + \mathbf{q} = \mathbf{k} \\ \mathbf{p,q} \in F } } {u}_j (\mathbf{p},t)  {u}_m (\mathbf{q},t) = - \bigg(\delta_{im} - \frac{k_i k_m}{k^2} \bigg) \imath k_j  \tau_{jm}(\mathbf{k},t) \qquad k \in F,
\end{equation}
where $F$ are the resolved modes and $G$ are the unresolved modes. Note that Eq.~\ref{eq:NSspec_filtered} is the LES equations one obtains if they apply a sharp spectral cutoff filter to the Navier-Stokes equations. The modes in $G$ are the unresolved modes that are filtered out, while the modes in $F$ are retained. The objective is to solve for the modes in $F$ as accurately as possible. The sub-grid stress is written as
$$\tau_{jm}(\mathbf{k},t) = \sum_{\substack{ \mathbf{p} + \mathbf{q} = \mathbf{k} \\ \mathbf{p,q} \in G } } {u}_j (\mathbf{p},t)  {u}_m (\mathbf{q},t) + 
 \sum_{\substack{ \mathbf{p} + \mathbf{q} = \mathbf{k} \\ \mathbf{p} \in G, \mathbf{q} \in F } } {u}_j (\mathbf{p},t)  {u}_m (\mathbf{q},t) + 
  \sum_{\substack{ \mathbf{p} + \mathbf{q} = \mathbf{k} \\ \mathbf{p} \in F, \mathbf{q} \in G } } {u}_j (\mathbf{p},t)  {u}_m (\mathbf{q},t).$$
Note that, in Fourier space, the pressure term appears as a projection. This projection leads to additional non-linear interactions between the resolved and unresolved scales. 

\subsection{Construction of the Mori-Zwanzig Models}
For the incompressible Navier-Stokes equations, the t-model and the first order finite memory model are considered. The expectation projection and Gaussian density in the zero variance limit are again used. Casually commuting the non-linear Markovian term, the projected Mori-Zwanzig identity reads
\begin{equation}\label{eq:NS_M-Z}
\bigg(\frac{\partial}{\partial t} + \nu k^2 \bigg){u}_{i} (\mathbf{k},t) + \bigg(\delta_{im} - \frac{k_i k_m}{k^2} \bigg) i k_j \sum_{\substack{ \mathbf{p} + \mathbf{q} = \mathbf{k} \\ \mathbf{p,q} \in F } } {u}_j (\mathbf{p},t)  {u}_m (\mathbf{q},t) = \MC{P}\int_0^t K(\mathbf{{u}}(t),t-s)ds  \qquad k \in F
\end{equation}
The evaluation of Eq.~\ref{eq:NS_M-Z} is made tractable by approximating the memory integral. Here only first order models are considered, which require the evaluation of $\MC{PLQL}u_{0k}$. After much tedious algebra it can be shown that
\begin{multline}\label{eq:NS_PLQLu}
e^{t \MC{L}}\MC{PLQL}u_{i}(\mathbf{k},0) = \bigg(-\delta_{im} +  \frac{k_i k_m}{k^2} \bigg)\imath k_j   \sum_{\substack{ \mathbf{p} + \mathbf{q} = \mathbf{k} \\ \mathbf{p} \in F, \mathbf{q} \in  G }} {u}_{j}(\mathbf{p},t)  e^{t \MC{L}}\MC{PL}u_{m}(\mathbf{q},0) -  \\
 \bigg(\delta_{im} + \frac{k_i k_m}{k^2} \bigg) \imath k_j\sum_{\substack{ \mathbf{p} + \mathbf{q} = \mathbf{k} \\ \mathbf{p} \in F, \mathbf{q} \in  G }} u_{m}(\mathbf{p},t) e^{tL} \MC{PL}u_{j}(\mathbf{q},0),
\end{multline}
where 
$$e^{t \MC{L}}\MC{PL}u_{i}(\mathbf{k},0) = \underbrace{-\nu k^2{u}_{i} (\mathbf{k},t)}_{k \in F} - \bigg(\delta_{im} - \frac{k_i k_m}{k^2} \bigg) \imath k_j \sum_{\substack{ \mathbf{p} + \mathbf{q} = \mathbf{k} \\ \mathbf{p,q} \in F } } {u}_j (\mathbf{p},t)  {u}_m (\mathbf{q},t) $$
Eq.~\ref{eq:NS_PLQLu}, along with Eqns.~\ref{eq:tmodel} and~\ref{eq:M-Z_FM1_2}, can be used to write equations for the t-model and the first order finite memory model. It is noted that the terms $\MC{PL}u_i$ are simply the right hand side of the filtered equations without any sub-grid model and are already computed. It is additionally noted that, when expanded, several terms in $\MC{PLQL}u_i$ can be combined for a faster evaluation.

\subsection{Numerical Implementation}
The Navier-Stokes equations are solved using a Galerkin spectral method with an explicit low storage RK4 time integration scheme.
The main solver is written in Python and interfaces with the FFTW discrete Fourier transform library~\cite{FFTW} via the pyFFTW wrapper. FFT calculations are padded by the 3/2 rule. For the Mori-Zwanzig models, a 2x padding is used such that $F \in [-N/2,N/2-1]$ and $G \in ([-N,-N/2-1],[N/2,N-1])$. Convolutions of the form
$$\sum_{\substack{ \mathbf{p} + \mathbf{q} = \mathbf{k} \\ \mathbf{p} \in F, \mathbf{q} \in  G }} {u}_{j}(\mathbf{p},t) {u}_{j}(\mathbf{q},t),$$ which have a support of $2N$, are padded by construction~\cite{stinisHighOrderEuler}.

\subsection{Homogeneous Isotropic Turbulence}
The simulation of decaying Homogeneous Isotropic Turbulence (HIT) is considered. HIT has been a baseline case for the development of sub-grid models. In this study, the spectrum used by Rogallo~\cite{Rogallo} is used for initialization. The velocity field is given by
\begin{equation}\label{eq:HITuIC}
u_i (\mathbf{k}) = \alpha e_i^1 + \beta e_i ^2,
\end{equation}
where $e_i^1$ and $e_i^2$ are mutually orthogonal unit vectors in the plane orthogonal to the wave vector. The initial spectrum is taken to be
$$E(k,0) = \frac{q^2}{2A} \frac{1}{k_p^{\sigma+1}} k^{\sigma} \exp \bigg(-\frac{\sigma}{2} \big(\frac{k}{k_p}\big)^2 \bigg),$$
where $k_p$ is the wave number at which the energy spectra is maximum, $\sigma$ is a parameter set to 4, and $A = \int_0^{\infty}k^{\sigma} \exp(-\sigma k^2/2)dk$.
DNS data from a $512^3$ simulation initialized with Eq.~\ref{eq:HITuIC} are used as an initial condition for LES simulations. The Taylor microscale-based Reynolds number of the filtered field is $Re_{\lambda} \approx 200.$ The LES simulations are performed using $64^3$ resolved modes with a time step of $\Delta t = 0.005$. The M-Z-based models are compared to filtered DNS data, both the dynamic and static Smagorinsky models, and an LES with no sub-grid model. An alternate heuristic  to select the memory length is to scale the time step in the LES with the ratio of the grid size to the estimated Kolmogorov scale. The relevant computational details are given in Table~\ref{table:HIT_summary}.

\begin{table}
\begin{center}
    \begin{tabular}{| l | l | l | l | l |}
    \hline
     HIT & DNS & Smagorinsky & t-model & Finite Memory  \\ \hline
    $N$ & $512^3$ & $64^3$ & $64^3$ & $64^3$ \\ \hline
$\Delta t$& 0.005 & 0.005 & 0.005 & 0.005 \\ \hline
Constants & NA & $Cs=0.16$ & NA & $\tau_0 = 0.1$ \\ \hline
    \end{tabular}
\end{center}
\caption{Summary of computational details for homogeneous isotropic turbulence case.}
\label{table:HIT_summary}
\end{table}

\begin{figure}
\begin{subfigure}[t]{1.\textwidth}
  \captionsetup{justification=justified,singlelinecheck=false}
  \includegraphics[trim={0.0cm 0 0cm 0cm},clip,width=.49\linewidth]{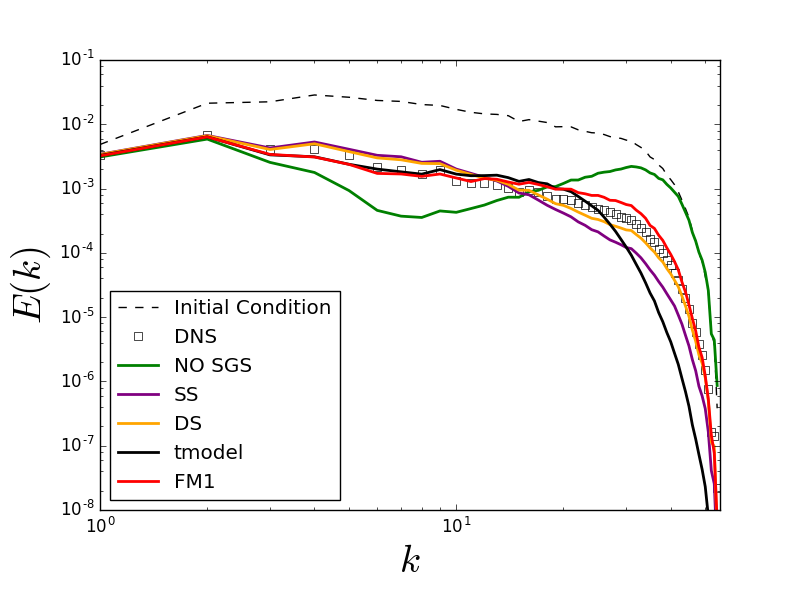}
  \includegraphics[trim={0.0cm 0 0cm 0cm},clip,width=.49\linewidth]{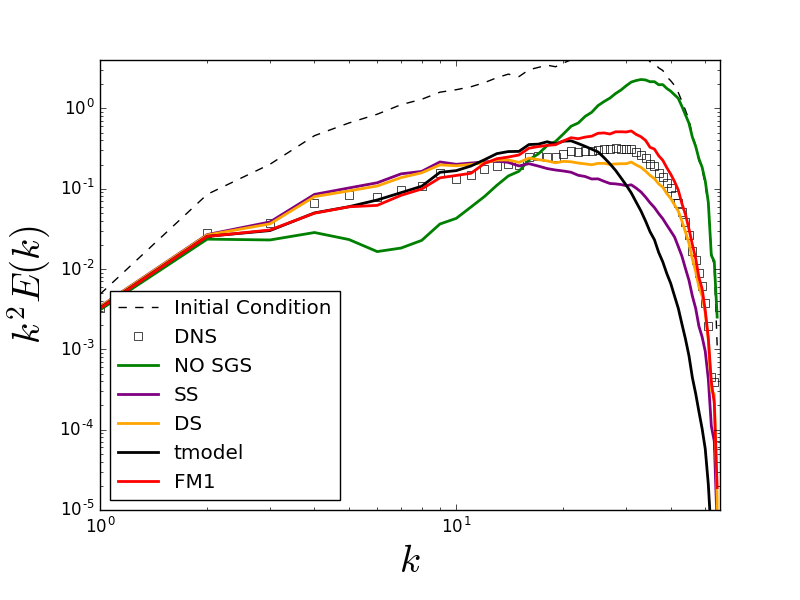}
  \caption{Filtered energy spectra (left) and filtered dissipation spectra (right).}
\end{subfigure}

\begin{subfigure}[t]{1.\textwidth}
  \captionsetup{justification=justified,singlelinecheck=false}
  \includegraphics[trim={0.0cm 0 0cm 0cm},clip,width=.49\linewidth]{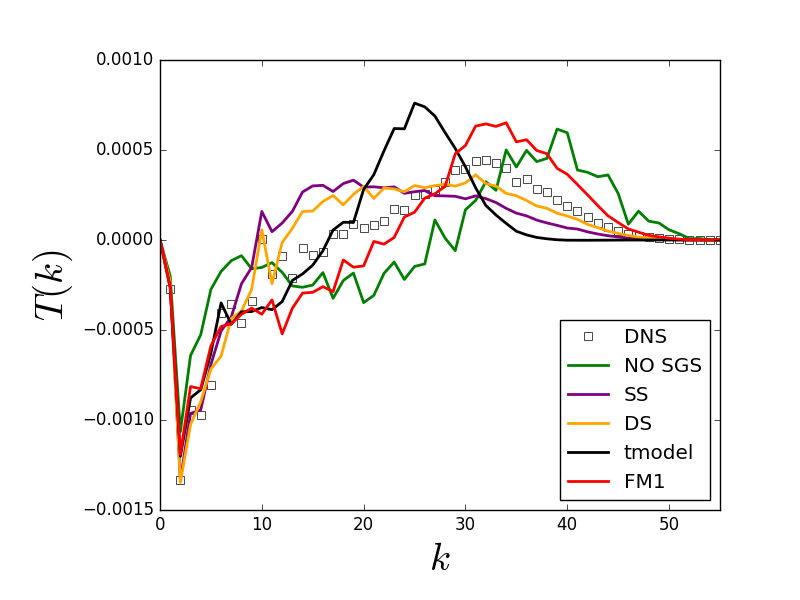}
  \includegraphics[trim={0.0cm 0 0cm 0cm},clip,width=.49\linewidth]{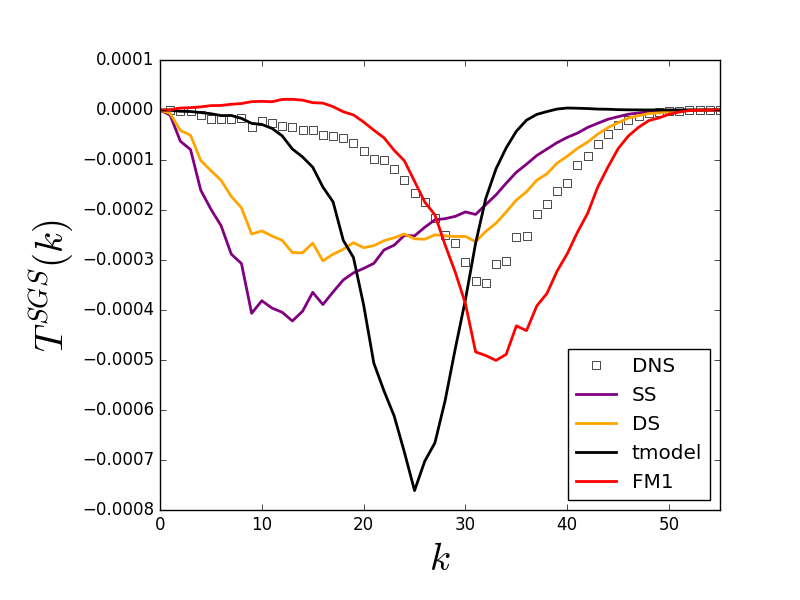}
  \caption{The transfer spectra as computed by the resolved modes are shown on the left. The sub-grid contribution to the transfer spectra from the DNS are compared to the contribution of the sub-grid models on the right.} 
\end{subfigure}

\caption{Energy, dissipation, and transfer spectra at $t = 4.0$ for the homogeneous isotropic turbulence case.} 
\label{fig:HIT_spectra} 
\end{figure}

Figure~\ref{fig:HIT_spectra} shows the energy, dissipation, resolved transfer spectra, and sub-grid transfer spectra at $t = 4.0$. Results are compared to filtered DNS data. The resolved transfer spectra is computed by
$$T(\mathbf{k}) =  -{u}^*_{i} (\mathbf{k},t) \bigg(\delta_{im} - \frac{k_i k_m}{k^2} \bigg) \imath k_j \sum_{\substack{ \mathbf{p} + \mathbf{q} = \mathbf{k} \\ \mathbf{p,q} \in F } } {u}_j (\mathbf{p},t)  {u}_m (\mathbf{q},t) \qquad \mathbf{k} \in F.
$$
The sub-grid energy transfer is extracted from the DNS data by
$$T^{SGS}(\mathbf{k}) = - {u}^*_{i} (\mathbf{k},t) \bigg(\delta_{im} - \frac{k_i k_m}{k^2} \bigg) \imath k_j \sum_{\substack{ \mathbf{p} + \mathbf{q} = \mathbf{k} \\ \mathbf{p,q} \notin F } } {u}_j (\mathbf{p},t)  {u}_m (\mathbf{q},t) \qquad \mathbf{k} \in F.
$$
For the large eddy simulations, the sub-grid energy transfer is computed by
$$T^{SGS}(\mathbf{k}) =  {u}^*_{i} (\mathbf{k},t) w_i^{(0)}(\mathbf{k},t)  \qquad \mathbf{k} \in F.$$
The simulation ran with no sub-grid model has a pileup of energy at high frequencies, indicating that the simulation is under-resolved. Both the static and dynamic Smagorinsky models provide good predictions for the energy, dissipation, and resolved spectra. The sub-grid-contribution to the energy spectra is qualitatively correct for both models, but error is present. 
The t-model performs well for low wave numbers, but under predicts the energy content at high wave numbers. 
The performance of the finite memory model is comparable to the dynamic Smagorinsky model and provides good predictions for all wave numbers. In particular, the finite memory model provides excellent predictions for the sub-grid-contributions to the transfer term. 

For the HIT case, it is perhaps not prudent to conclude that the M-Z-based models 
performed significantly better  than the Smagorinsky models. 
The derivation of the Smagorinsky models, however, proceeds directly from equilibrium assumptions that are most relevant specifically in homogeneous isotropic turbulence. Further, there is an implicit assumption that 
the simulation has resolution into the inertial subrange. 
The M-Z models, on the other hand, are insensitive to any 
assumptions about the state of the flow. 
This generality allows the models to be used in a variety of flow regimes, including ones where use of the Smagorinsky model is inappropriate. This robustness is evident in the next example, where the Taylor Green vortex is considered. 

\subsection{Taylor Green Vortex}
The Taylor Green Vortex (TGV) is a canonical problem with deterministic initial conditions that is often used to study the accuracy of computational methods. The flow is characterized by a breakdown of organized 
coherent structures into fine-scale features. The initial conditions are
\begin{align*}
u_1(x_1,x_2,x_3) =& \hspace{0.02in}\cos(x_1)\sin(x_2)\cos(x_3) \\
u_2(x_1,x_2,x_3) =& \hspace{0.02in}-\sin(x_1)\cos(x_2)\cos(x_3) \\
u_3(x_1,x_2,x_3) =& \hspace{0.02in} 0,
\end{align*}
in a periodic domain $x_1,x_2,x_3 \in [-\pi,\pi]$. The Reynolds number of the flow is given by the inverse of viscosity. Traditional LES sub-grid models do not perform well on the TGV since they are generally designed for fully developed turbulent flows. 
Two cases are considered, one at $Re = 800$ and the other at $Re=1600$. All comparisons are made to filtered DNS quantities. Tables~\ref{table:TGV_800_summary} and~\ref{table:TGV_1600_summary} summarize the relevant simulation details. Note that the t-model required a lower time step for stability.
\begin{table}
\begin{center}
    \begin{tabular}{| l | l | l | l | l |}
    \hline
     Re=800 & DNS & Smagorinsky & t-model & Finite Memory  \\ \hline
    $N$ & $128^3$ & $32^3$ & $32^3$ & $32^3$ \\ \hline
$\Delta t$& 0.005 & 0.02 & 0.005 & 0.02 \\ \hline
Constants & NA & $Cs=0.16$ & NA & $\tau_0 = 0.1$ \\ \hline
    \end{tabular}
\end{center}
\caption{Summary of computational details for Taylor Green Vortex cases for Re=800.}
\label{table:TGV_800_summary}
\end{table}

\begin{table}
\begin{center}
    \begin{tabular}{| l | l | l | l | l |}
    \hline
     Re=1600 & DNS & Smagorinsky & t-model & Finite Memory  \\ \hline
    $N$ & $256^3$ & $32^3$ & $32^3$ & $32^3$ \\ \hline
$\Delta t$& 0.005 & 0.02 & 0.005 & 0.02 \\ \hline
Constants & NA & $Cs=0.16$ & NA & $\tau_0 = 0.1$ \\ \hline
    \end{tabular}
\end{center}
\caption{Summary of computational details for Taylor Green Vortex cases for Re=1600.}
\label{table:TGV_1600_summary}
\end{table}

\begin{figure}
\begin{subfigure}[t]{1.\textwidth}
  \captionsetup{justification=justified,singlelinecheck=false}
  \includegraphics[trim={0.05cm 0 1.75cm 1cm},clip,width=.49\linewidth]  {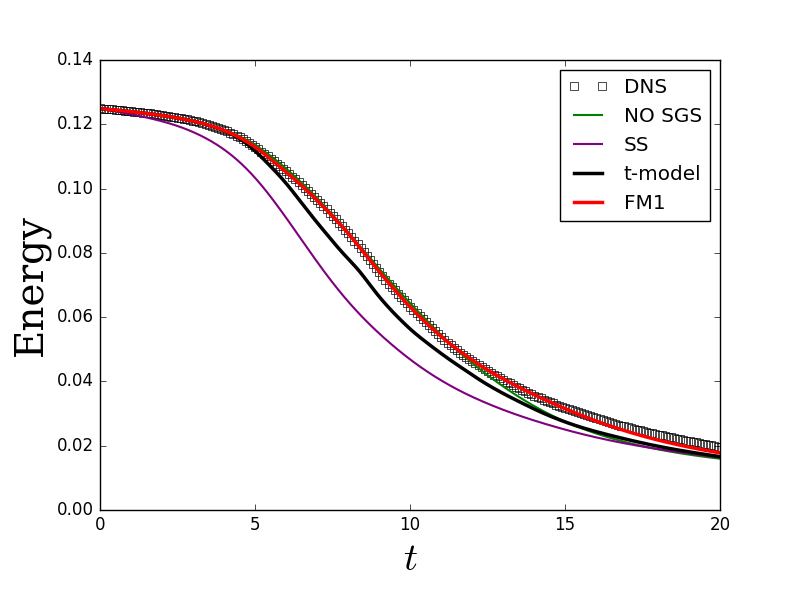}
  \includegraphics[trim={0.05cm 0 1.75cm 1cm},clip,width=.49\linewidth]  {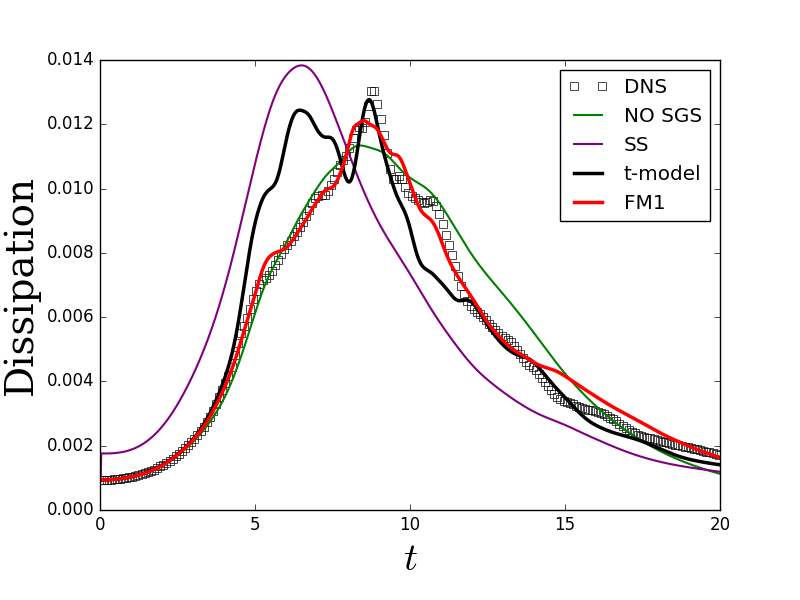}
  \caption{Evolution of integral quantities for Re=800 case.}
  \label{fig:TGV_800_ED}
\end{subfigure}

\begin{subfigure}[t]{1.\textwidth}
  \captionsetup{justification=justified,singlelinecheck=false}
  \includegraphics[trim={0.05cm 0 1.75cm 1cm},clip,width=.49\linewidth]{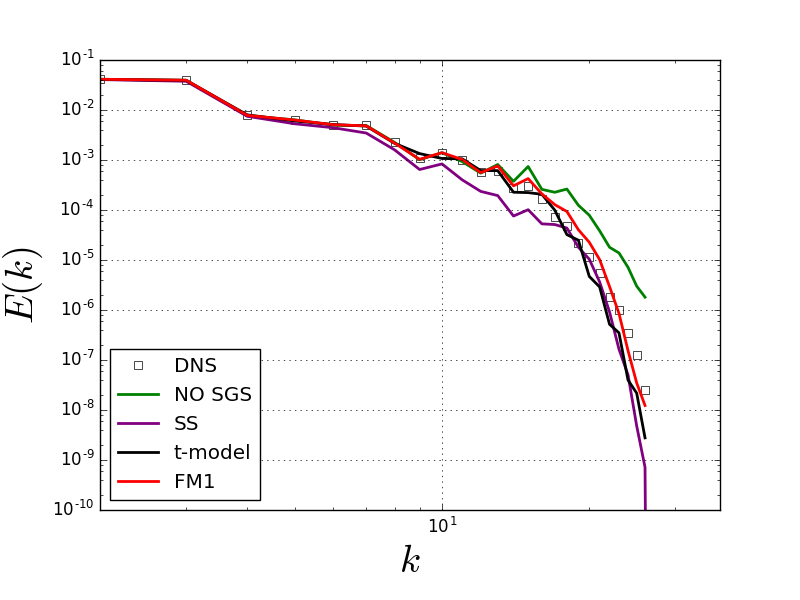}
  \includegraphics[trim={0.05cm 0 1.75cm 1cm},clip,width=.49\linewidth]{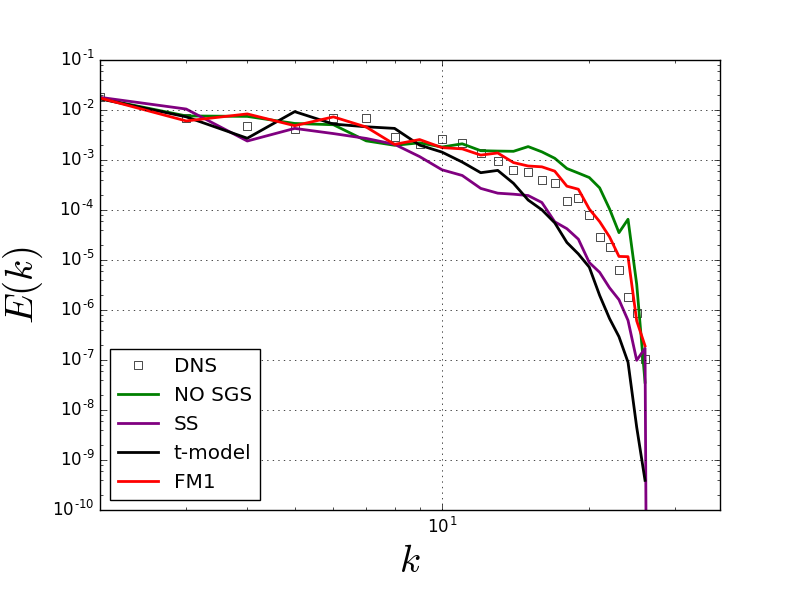}
  \caption{Energy spectra at $t=5$ (left) and $t=10$ (right) for Re=800 case. }
  \label{fig:TGV_800_Espec}
\end{subfigure}

\begin{subfigure}[t]{1.\textwidth}
  \captionsetup{justification=justified,singlelinecheck=false}
  \includegraphics[trim={0.05cm 0 1.75cm 1cm},clip,width=.49\linewidth]{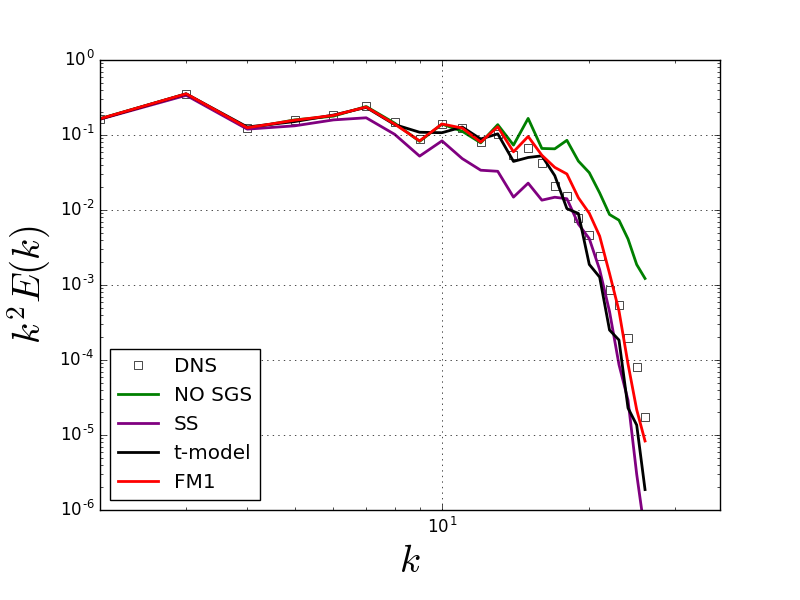}
  \includegraphics[trim={0.05cm 0 1.75cm 1cm},clip,width=.49\linewidth]{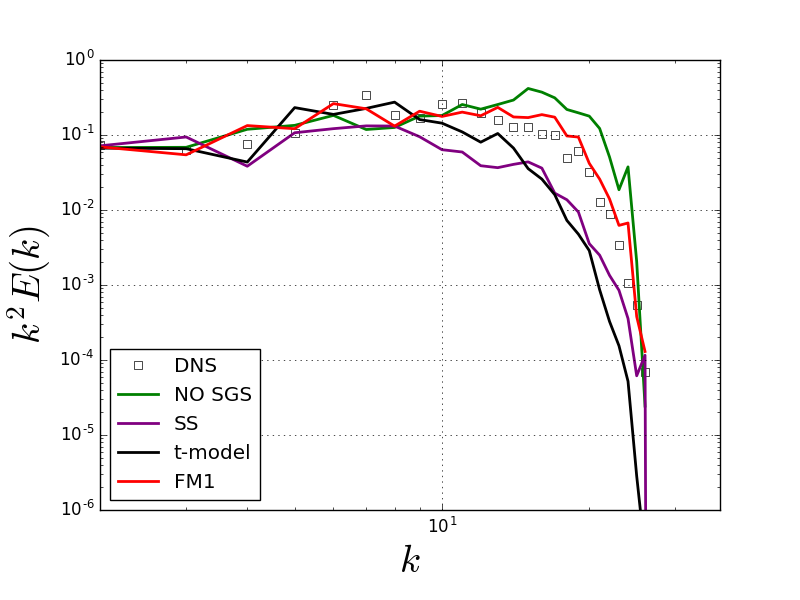}
  \caption{Dissipation spectra at $t=5$ (left) and $t=10$ (right) for Re=800 case. }
  \label{fig:TGV_800_Dspec}
\end{subfigure}
\caption{ Results for numerical simulations of the Taylor Green Vortex at ${Re=800}$. DNS quantities are obtained from filtered data obtained on a ${128^3}$ grid. All other models are ran on ${32^3}$ grids.} 
\label{fig:TGV_800}
\end{figure}

The results of the $Re=800$ case are shown in Figure~\ref{fig:TGV_800}. The large eddy simulations are performed with $32^3$ resolved modes, while the DNS simulation uses $128^3$ resolved modes. At this Reynolds number and resolution, the system is only slightly under-resolved, as evidenced by the reasonable performance of the simulation with no sub-grid model. Figure~\ref{fig:TGV_800_ED} shows the temporal evolution of the kinetic energy integrated over the whole domain as well as the temporal evolution of the kinetic energy dissipation rate (computed by $-\frac{dE}{dt}$). The performance of the Smagorinsky model is poor. As was previously discussed, this is due to the fact that the model is designed for fully turbulent flows. The Smagorinsky model is unable to account for the fact that the solution is fully resolved at early time, and incorrectly removes energy from the resolved scales at $t=0$. The Mori-Zwanzig models, particularly the finite memory model, perform notably better. The t-model recognizes that the simulation is completely resolved at early times, but is seen to remove too much energy from the resolved modes soon after. The finite memory model provides good predictions for the total energy and dissipation. In particular, the model is able to capture the double peaked structure of the kinetic energy dissipation rate around $t=8$. The energy and dissipation spectra of the simulations at $t=5$ and $t=10$ are shown in Figures~\ref{fig:TGV_800_Espec} and \ref{fig:TGV_800_Dspec}. The finite memory model is again in good agreement with the DNS. The spectra predicted by the t-model is in good agreement with the DNS for early time, but the dissipative nature of the model is evident at $t=10$. 

\begin{figure}
\begin{subfigure}[t]{1.\textwidth}
  \captionsetup{justification=justified,singlelinecheck=false}
  \includegraphics[trim={0.05cm 0 1.75cm 1cm},clip,width=.49\linewidth]  {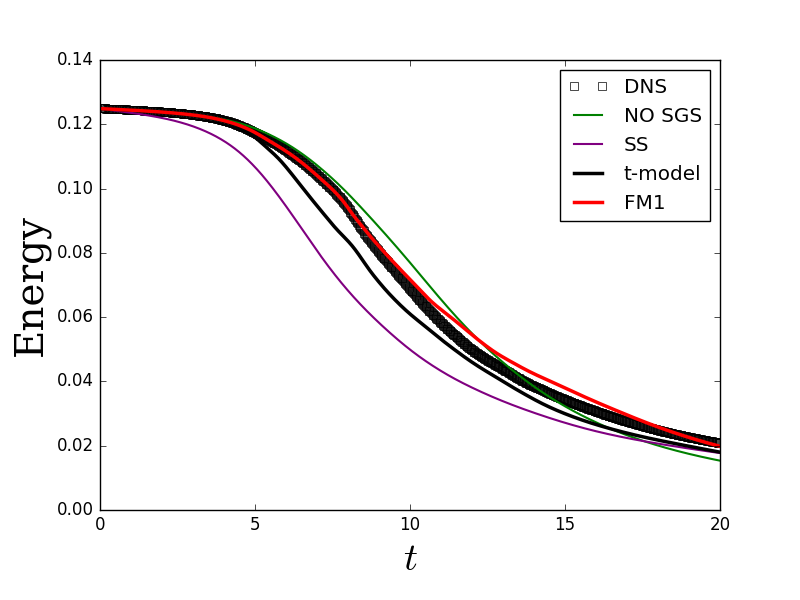}
  \includegraphics[trim={0.05cm 0 1.75cm 1cm},clip,width=.49\linewidth]  {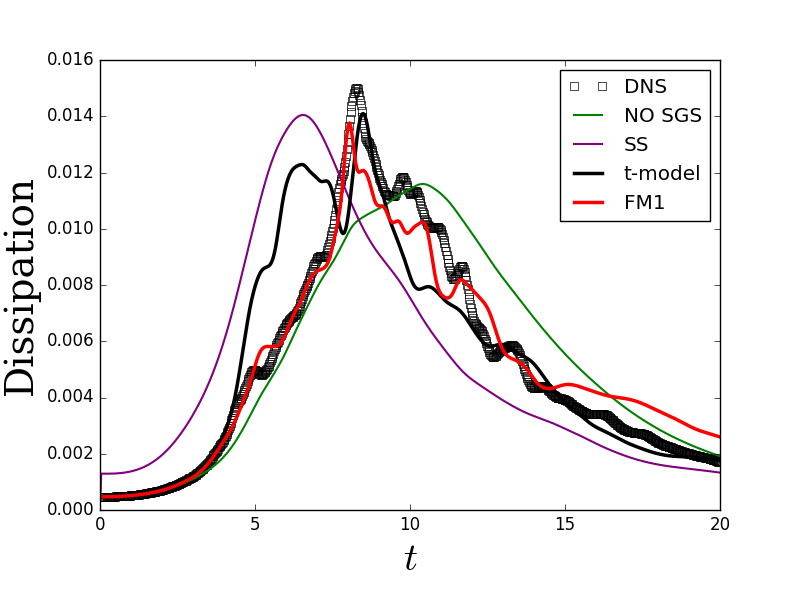}
  \caption{Evolution of integral quantities for Re=1600 case.  }
  \label{fig:TGV_1600_ED}
\end{subfigure}

\begin{subfigure}[t]{1.\textwidth}
  \captionsetup{justification=justified,singlelinecheck=false}
  \includegraphics[trim={0.05cm 0 1.75cm 1cm},clip,width=.49\linewidth]{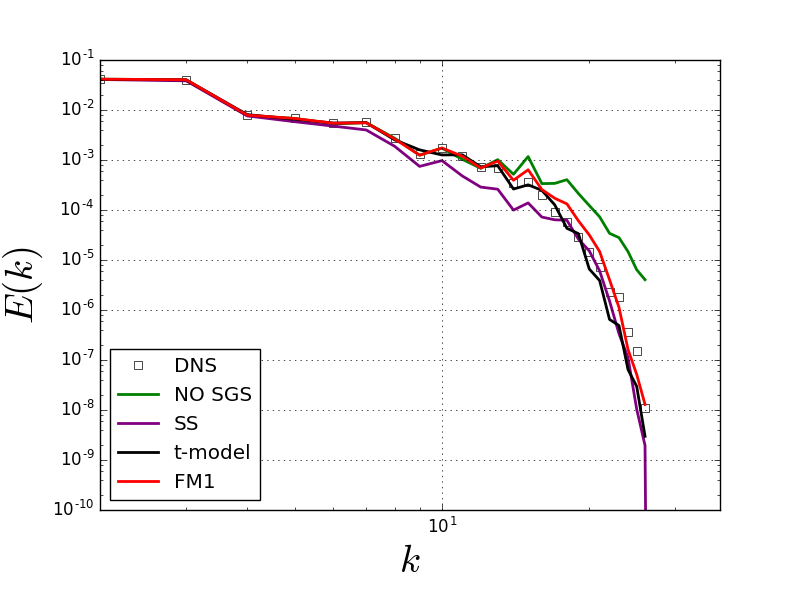}
  \includegraphics[trim={0.05cm 0 1.75cm 1cm},clip,width=.49\linewidth]{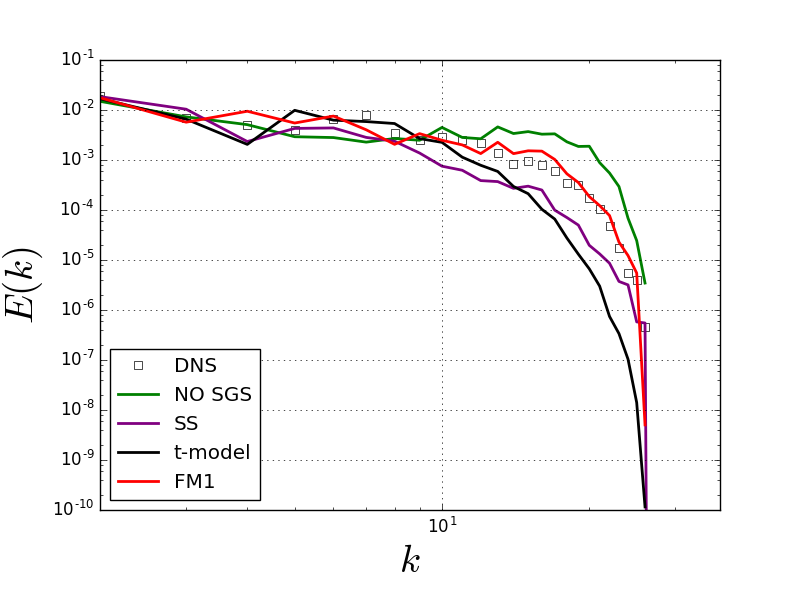}
  \caption{Energy spectra at $t=5$ (left) and $t=10$ (right) for Re=1600 case.}
  \label{fig:TGV_1600_Espec}
\end{subfigure}

\begin{subfigure}[t]{1.\textwidth}
  \captionsetup{justification=justified,singlelinecheck=false}
  \includegraphics[trim={0.05cm 0 1.75cm 1cm},clip,width=.49\linewidth]{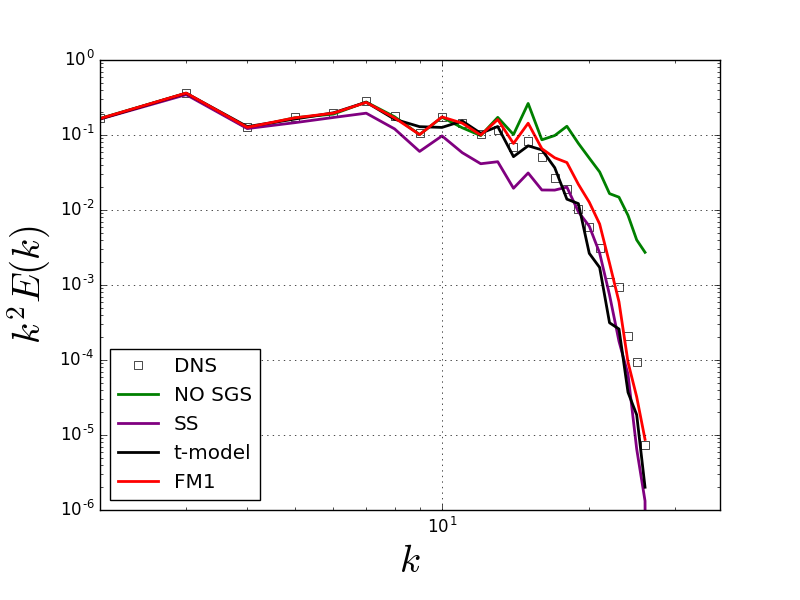}
  \includegraphics[trim={0.05cm 0 1.75cm 1cm},clip,width=.49\linewidth]{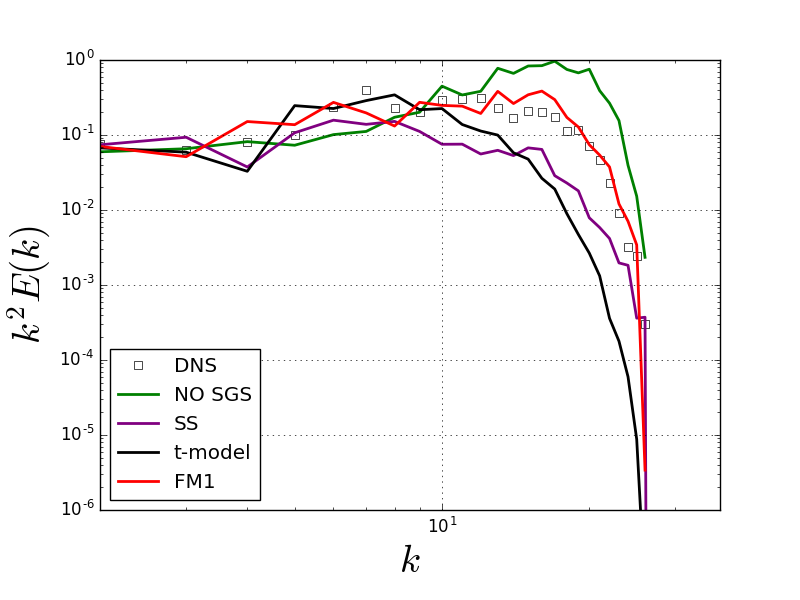}
  \caption{Dissipation spectra at $t=5,$ (left) and $t=10$ (right) for Re=1600 case. }
  \label{fig:TGV_1600_Dspec}
\end{subfigure}

\caption{ Results for numerical simulations of the Taylor Green Vortex at ${Re=1600}$. DNS quantities are obtained from filtered data obtained on a ${256^3}$ grid. All other models are ran on ${32^3}$ grids.} 
\label{fig:TGV_1600}
\end{figure}

The results of the $Re=1600$ case are shown in Figure~\ref{fig:TGV_1600}. The LES is again performed with $32^3$ resolved modes, while the DNS  used $256^3$  modes. The sub-grid models in the coarse-grained simulations are more active at this higher Reynolds number. The  finite memory model performs well. The temporal evolution of the total kinetic energy and dissipation of kinetic energy are well characterized, showing that the model is removing energy from the resolved scales at an appropriate rate. The predicted spectra are also in good agreement. The t-model and Smagorinsky models are again highly dissipative. 


%

\section{Application to Fully Developed Channel Flow}
An M-Z-based finite memory model is now applied to turbulent channel flow. Unlike the previously considered cases, the channel flow is a steady-state non-decaying problem. For such problems, a finite memory assumption is critical for the application of M-Z-based models. The construction of a first order finite memory M-Z-based model for fully developed turbulent channel flow is now outlined. The flow is taken to be streamwise (x) and spanwize (z) periodic. The model is constructed by coarse-graining in the periodic directions.
\subsection{Construction of the Finite Memory Mori-Zwanzig Model}
Fourier transforming the incompressible Navier-Stokes equations in the $x$ and $z$ direction yields 
\begin{equation}\label{eq:3D-cont}
\frac{\hat{\partial}}{\partial \hat{x}_j} \hat{u}_j(\mathbf{k},t) = 0
 \end{equation}
\begin{equation}\label{eq:3Dmomentum-x}
\frac{\partial }{\partial t} \hat{u}_i( \mathbf{k},t) + \frac{\hat{\partial}}{\partial \hat{x}_j}  \sum_{\substack{ \mathbf{p} + \mathbf{q} = \mathbf{k} \\ \mathbf{p} ,\mathbf{q} \in F \cup G }}\hat{u}_i( \mathbf{p} ,t)  \hat{u}_j (\mathbf{q},t)
= - \frac{1}{\rho} \frac{\hat{\partial}}{\partial \hat{x}_i} \hat{p}( \mathbf{k},t) + \nu \bigg(-k_1^2 - k_3^2  + \frac{\partial^2}{\partial y^2} \bigg) \hat{u}_i (\mathbf{k},t)
\end{equation}
where 
$$\frac{\hat{\partial}}{{\partial \hat{x}_j}} = \big\{ \imath k_1, \partial_y, \imath k_3 \big\}.$$
Unlike in the triply periodic case, the continuity equation can not be implicitly satisfied by a simple solution to the pressure Poisson equation in Fourier space. Solution of the pressure Poisson equation is complicated by inhomogeneity in the $y$ direction and boundary conditions. This makes the derivation of M-Z models difficult as the framework is formulated for time varying dynamical systems. However, the effect of the pressure projection on the sub-grid scale models for triply periodic problems has been observed to be minimal. As such, the M-Z models are formulated by neglecting the effects induced by coarse-graining the pressure. In this case one finds
\begin{equation}
\MC{PLQL}\hat{u}_{i}(\mathbf{k}) =  - \frac{\hat{\partial}}{\partial \hat{x}_j} \sum_{\substack{ p + q = \mathbf{k} \\ p \in F ,q \notin F }} \hat{u}_j (\mathbf{p}) \MC{PL}\hat{u}_i (\mathbf{q})   - 
\frac{\hat{\partial}}{\partial \hat{x}_j} \sum_{\substack{ p + q = \mathbf{k} \\ p \in F ,q \notin F }} \hat{u}_i (\mathbf{p}) \MC{PL}\hat{u}_j (\mathbf{q} ) , \qquad \mathbf{k} \in F  .
\end{equation}

\subsection{Numerical Implementation}
The Navier-Stokes equations are solved in skew-symmetric form via a Fourier-Chebyshev pseudo-spectral method. A coupled semi-implicit Adams-Bashforth  scheme is used for time integration, as in~\cite{MoinChannel1}. The continuity equation is directly enforced at each time-step, bypassing the need for pressure boundary conditions. The main solvers are written in Python and utilize mpi4py for parallelization. All FFT calculations (including the Chebyshev transforms) are de-aliased by the 3/2 rule.

\subsection{Numerical Results}
Large Eddy Simulations of channel flow are now discussed. The solutions are compared to the dynamic Smagorinsky model. 
Simulations at $Re_{\tau} = 180$ are considered. The LES simulations are evolved using $32 \times 64 \times 32$ resolved modes in the $x,y,$ and $z$ directions respectively. Simulation parameters are given in Table~\ref{table:2}.
\begin{table}
\begin{center}\scriptsize
\vskip -0.1in
\begin{tabular}{c c c c c c c c}
\hline
 $L_x$ & $L_y$ & $L_z$ & $Re_{\tau}$ & $N_x$  & $N_y$ &$N_z$ & $\Delta t$ \\
 4 $\pi$ & 2 & $2 \pi $ & 180 & 32 & 64 & 32  & 0.01 \\
 \hline
\end{tabular}
\caption{Physical and numerical details for Large Eddy Simulations of the channel flow.}
\label{table:2}
\end{center}
\end{table}

\begin{figure}
\centering
\includegraphics[width=0.95\textwidth]{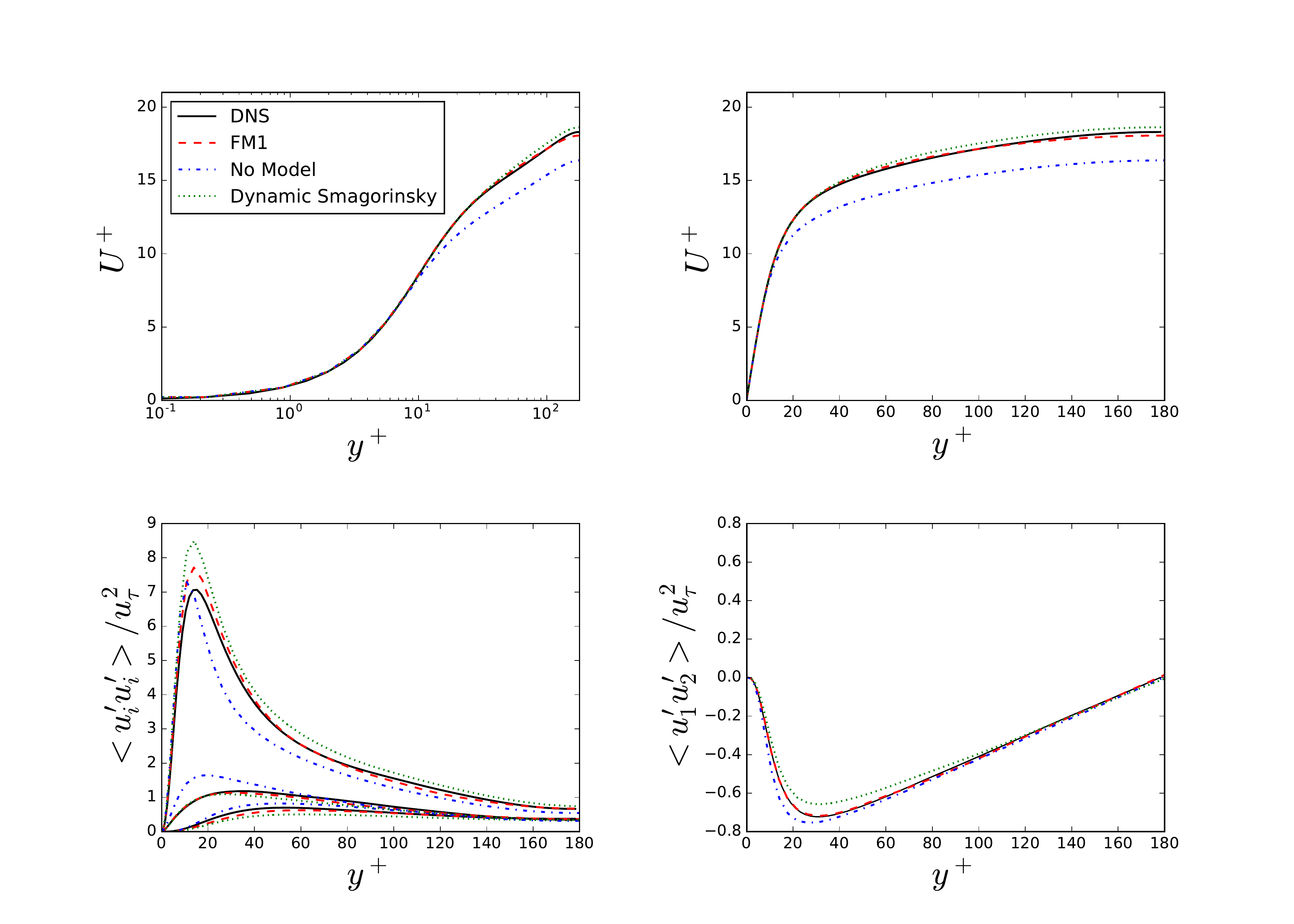}
	\caption{Statistical properties for fully developed channel flow at $Re_{\tau} = 180$.}
	\label{fig:Channel1}
\end{figure}

Statistical properties of the LES solutions are compared to DNS data from~\cite{MoserChannel} in Figure~\ref{fig:Channel1}. 
The M-Z-based models are seen to offer improved solutions similar to that produced from the dynamic Smagorinsky model. In particular, the mean velocity profiles are much improved and the model correctly reduces the Reynolds stresses. 

\section{Conclusions and Perspectives}
The Mori-Zwanzig formalism provides a mathematically consistent framework for model order reduction. Recasting a high-order dynamical system into the generalized Langevin equation (GLE) provides both a starting point for the development of coarse-grained models as well as insight into the effects of coarse-graining. Utilizing insight gained from solutions of the orthogonal dynamics equation for linear dynamical systems, a class of models based on the assumption that the memory convolution integral in the GLE has a finite, time-dependent support (Stinis, 2012) was presented. The appeal of these models is that they are formally derived from the governing equations and require minimal heuristic arguments.

Coarse-grained simulations of the viscous Burgers equation and the incompressible Navier-Stokes equations were considered. The closures derived under the assumption of a finite memory proved to be robust and accurate for the cases considered. For varying mesh resolutions, the M-Z-based models were shown to accurately predict the sub-grid contribution to the energy transfer. The trajectory of each resolved mode in phase-space was accurately predicted for cases where the coarse-graining was moderate. The models provided accurate results for Burgers equation, transitional and fully turbulent periodic incompressible flows, and fully developed channel flow. This accuracy is believed to be due to the close link between the mathematics of the coarse-graining process and the derivation of the closure model. An analysis of the memory term for Burgers equation demonstrated the need for the finite memory length and provided further insight into the performance of the t-model and finite memory models. The models used in this work should be considered a first order approximation. Extensions to the models, such as a spatio-temporal memory lengths, models for the unclosed $w_j^{(n+1)}$ terms, etc., are yet to be considered. 

The models are capable of addressing non-local effects characteristic of systems that lack scale-separation. The only heuristic that was used in the model development is an estimation of the length of the memory ({\em i.e.} the time scale of the convolution memory integral) based on the spectral radius of the Jacobian of the resolved variables. An alternate heuristic is the scaling of the time step in the LES with the ratio of the grid size to the estimated Kolmogorov scale. We note that more rigorous methodologies such as renormalization~\cite{RenormalizedMZ} and dynamical procedures~\cite{Parish_Dtau} can be used to derive estimates of the memory length. In this work, our objective was to mainly gain insight into the mechanics of the memory kernel.


 M-Z-based approaches have not gained substantial exposure in the fluid mechanics community and the results presented in this work highlight the promise of these techniques as a basis for LES.  Extension to practical problems requires further development. In particular, the Mori-Zwanzig procedure requires that one can discretize the governing equations as a dynamical system that can be naturally separated into resolved and unresolved sets. The Fourier-Galerkin approach, as was used in this work, is perhaps the most natural setting for such a hierarchical description. The Fourier-Galerkin approximation, however, restricts the applicability to periodic problems. The extension to traditional finite element/volume/difference schemes will require a formulation that makes use of specialized scale-separation operators. The variational multiscale method~\cite{gravemeir,hughes} and spectral element method are two such promising candidates.  

The models considered in this work were derived under the assumption of a Gaussian density 
in the zero-variance limit for the initial conditions. For problems that involve initial conditions that are unresolved, it 
will be appropriate to derive the models under a non-zero variance. 
Further investigations are required regarding the stability of the resulting models. 
Tests confirmed that the t-model required a lower time-step for stability compared to the finite memory models. 
Future work should explore opportunities to enforce physical constraints such as Galilean invariance of the sub-grid stress. 

\section{Acknowledgments}
This research was supported by the National Science Foundation via grant 1507928 (Technical monitor: Ron Joslin).  Computing resources were provided by the NSF via grant {\em MRI: Acquisition of Conflux, A Novel Platform for Data-Driven Computational Physics} (Tech. Monitor: Ed Walker).

\begin{appendices}
\section{Higher Order Models for the Burgers Equation}\label{sec:appendix0}
The higher order finite memory models FM2 and FM3 require the evaluation of $\MC{PLQLQL}u_{0k}$ and $\MC{PLQLQLQL}u_{0k}$. The forms follow a pattern that is similar to Pascal's triangle and can be verified to be
\begin{multline}e^{t \MC{L}}\MC{PLQLQL}u_{0k} = 2 \bigg(\frac{-\imath k}{2} \sum_{ p + q =k} u_p  \big[e^{t \MC{L}}\MC{PLQL}{u_{0k}}\big]_q  \bigg) + 2 \bigg(\frac{-\imath k}{2} \sum_{ p + q =k} \big[e^{t \MC{L}}\MC{PL}u_{0k} \big]_p \big[e^{t \MC{L}}\MC{PL}{u_{0k}}\big]_q  \bigg) + \\ 2 \bigg(\frac{-\imath k}{2} \sum_{ p + q =k } \big[e^{t \MC{L}}\MC{PL}u_{0k} \big]_q \big[e^{t \MC{L}}\MC{PL}{u_{0k}}\big]_q  \bigg) - \nu k^2 \big[e^{t \MC{L}}\MC{PLQL}u_{0k}\big]_q
\end{multline}
\begin{multline}e^{t \MC{L}}\MC{PLQLQLQL}u_{0k} = 2 \bigg(\frac{-\imath k}{2} \sum_{ p + q =k} u_p  \big[e^{t \MC{L}}\MC{PLQLQL}{u_{0k}}\big]_q  \bigg) + \\ 4 \bigg(\frac{-\imath k}{2} \sum_{ p + q =k} \big[e^{t \MC{L}}\MC{PL}u_{0k} \big]_p \big[e^{t \MC{L}}\MC{PLQL}{u_{0k}}\big]_q  \bigg) +  
4 \bigg(\frac{-\imath k}{2} \sum_{ p + q =k} \big[e^{t \MC{L}}\MC{PLQL}u_{0k} \big]_p \big[e^{t \MC{L}}\MC{PL}{u_{0k}}\big]_q  \bigg) +  \\
2 \bigg(\frac{-\imath k}{2} \sum_{ p + q =k} \big[e^{t \MC{L}}\MC{PLPL}u_{0k} \big]_p \big[e^{t \MC{L}}\MC{PL}{u_{0k}}\big]_q  \bigg) + 
  6 \bigg(\frac{-\imath k}{2} \sum_{ p + q =k } \big[e^{t \MC{L}}\MC{PLQL}u_{0k} \big]_q \big[e^{t \MC{L}}\MC{PL}{u_{0k}}\big]_q  \bigg) \\ + 2 \bigg(\frac{-\imath k}{2} \sum_{ p + q =k } \big[e^{t \MC{L}}\MC{PLPL}u_{0k} \big]_q \big[e^{t \MC{L}}\MC{PL}{u_{0k}}\big]_q  \bigg) - \nu k^2 \big[e^{t \MC{L}}\MC{PLQLQL}u_{0k}\big]_q.
\end{multline}
\section{Burgers equation: Sensitivity of Results to Memory Length}\label{sec:appendix1}
The finite memory models discussed require the specification of a memory length. In Section~\ref{sec:selectMemoryLength}, an empirical methodology to determine the memory length was presented. It was shown that the optimal memory constant $\tau_0$ (in the $L^2$ sense of the error in predicted dissipation rate) can be approximated by a linear scaling of the spectral radius of the Jacobian. The reader will note that the collapse of the data, although good, was not perfect. This scatter gives rise to a statistical uncertainty in the memory length and raises concern over the dependence of the model results on the memory length. To provide a measure of this uncertainty, we consider a statistical parametrization of the inferred memory constant for the viscous Burgers equation.

\begin{figure}
\begin{centering}
\includegraphics[trim={0.05cm 0 1.75cm 1cm},clip,width=.49\linewidth]  {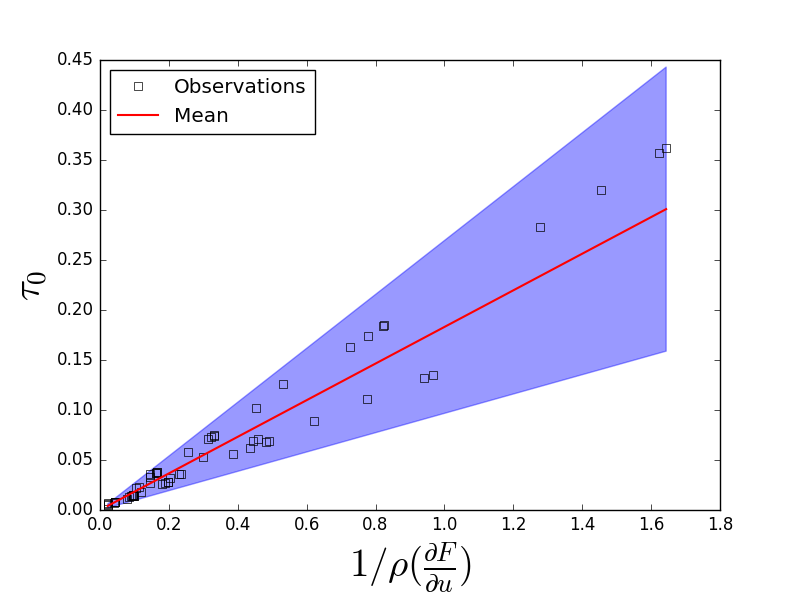}
\caption{Modeled probability distribution for $\tau_0$. The mean of the Gaussian PDF is shown in red. The shaded regions correspond to the 95\% confidence intervals. } 
\label{fig:tau_distribution}
\end{centering}
\end{figure}

As in Section~\ref{sec:selectMemoryLength}, assume that the memory constant is given by the linear model
$$\tau_0 = w  \bigg[ \rho \bigg(\frac{\partial F}{\partial u}\bigg) \bigg]^{-1}$$
where $w$ is the weighting parameter.
The regression problem is now approached from a stochastic point of view, where $\tau$ (and hence $w$) is assumed to be a random variable. We assume $w$ to be Gaussian with mean $\mu_0$ and variance $\sigma^2$. To determine the mean and variance of the distribution, linear regression is performed for each of the data points inferred in Section~\ref{sec:selectMemoryLength}. The results of the regression provide a distribution for the weight $w$ from which we compute a mean and variance. This process minimizes over-fitting the data. The resulting model distribution is shown in Figure~\ref{fig:tau_distribution}. Note that approximating the distribution for $w$ as Gaussian is not particularly accurate since the inferred results for $\tau_0$ are bi-modal. However, the linear model itself is an approximation and the Gaussian distribution is sufficient to demonstrate the sensitivity of the models to the memory length.

\begin{figure}
\begin{subfigure}[t]{1.\textwidth}
  \captionsetup{justification=justified,singlelinecheck=false}
  \includegraphics[trim={0.05cm 0 1.75cm 1cm},clip,width=.32\linewidth]{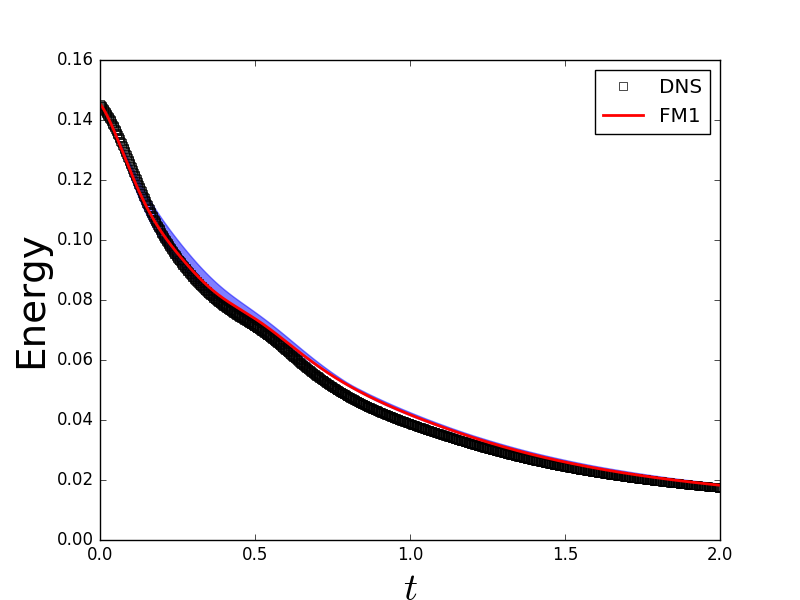}
  \includegraphics[trim={0.05cm 0 1.75cm 1cm},clip,width=.32\linewidth]{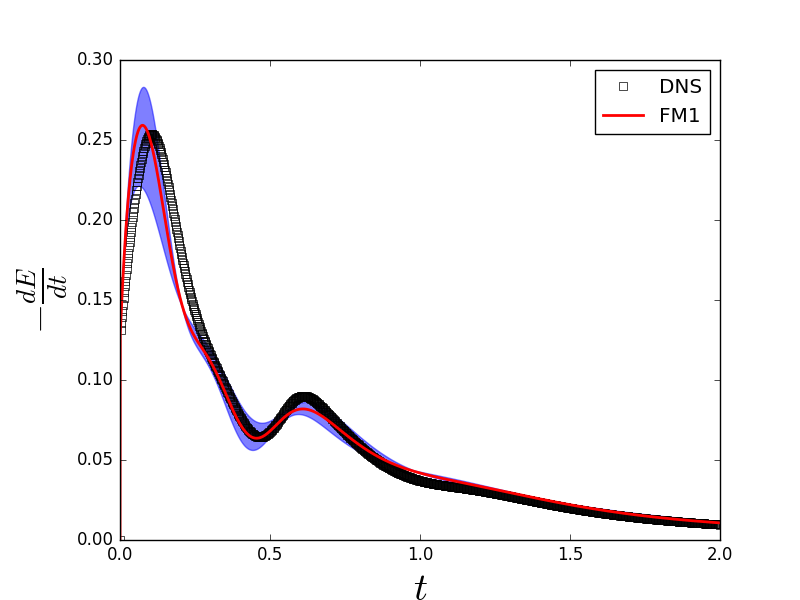}
  \includegraphics[trim={0.05cm 0 1.75cm 1cm},clip,width=.32\linewidth]{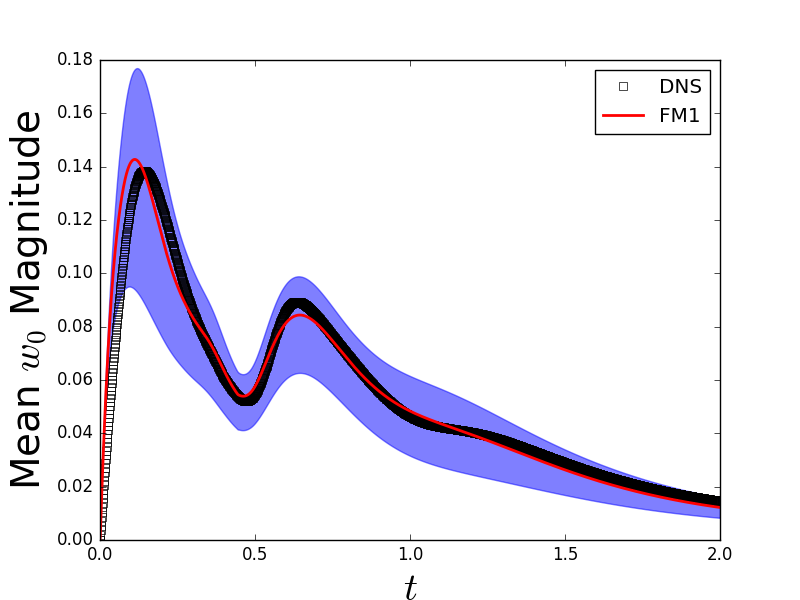}
  \caption{Probability distributions for the VBE case using $k_c=16,\nu=0.01$, and $U_0^* = 1.$}
  \label{fig:burg_post1}
\end{subfigure}

\begin{subfigure}[t]{1.\textwidth}
  \captionsetup{justification=justified,singlelinecheck=false}
  \includegraphics[trim={0.05cm 0 1.75cm 1cm},clip,width=.32\linewidth]{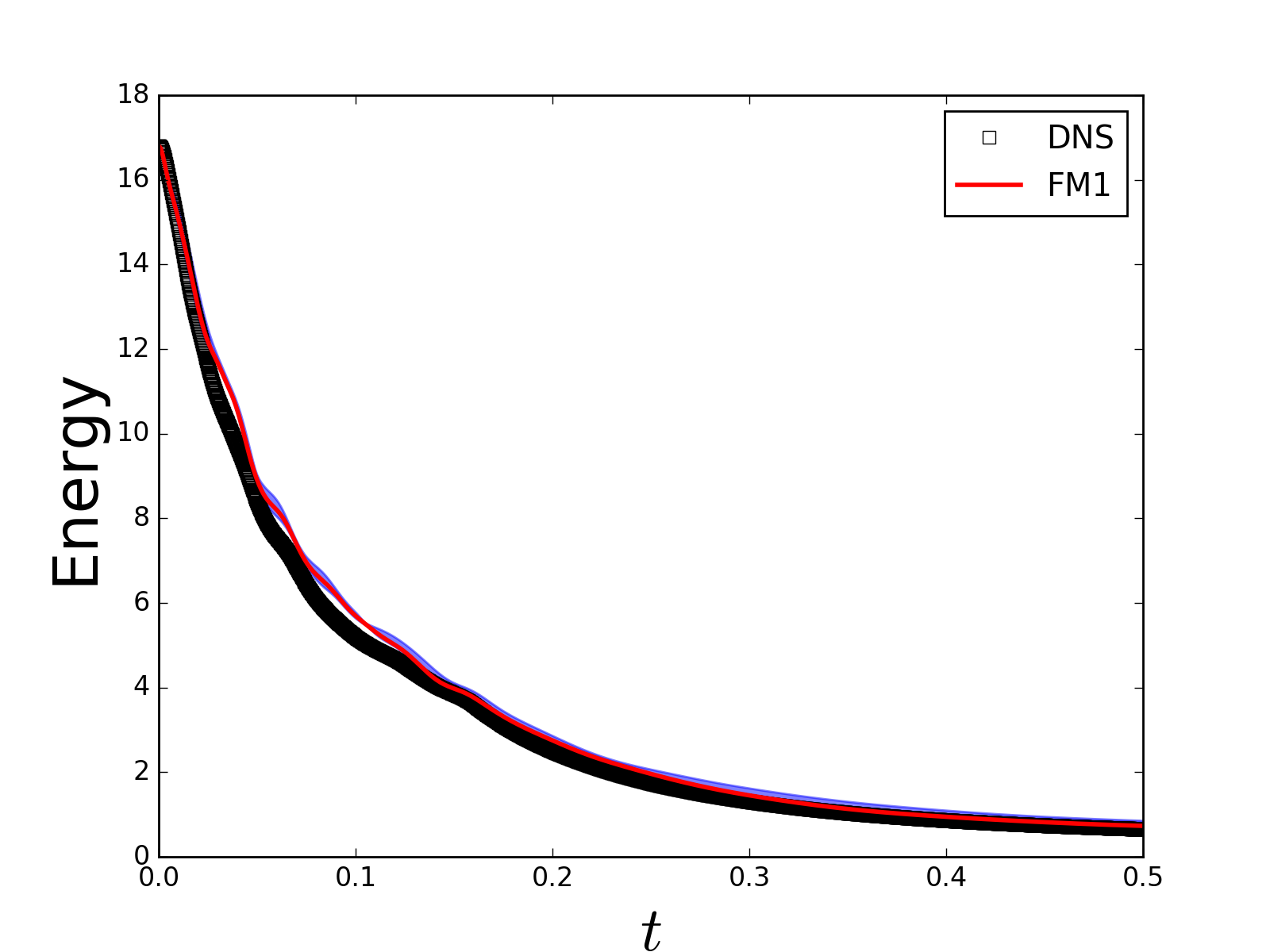}
  \includegraphics[trim={0.05cm 0 1.75cm 1cm},clip,width=.32\linewidth]{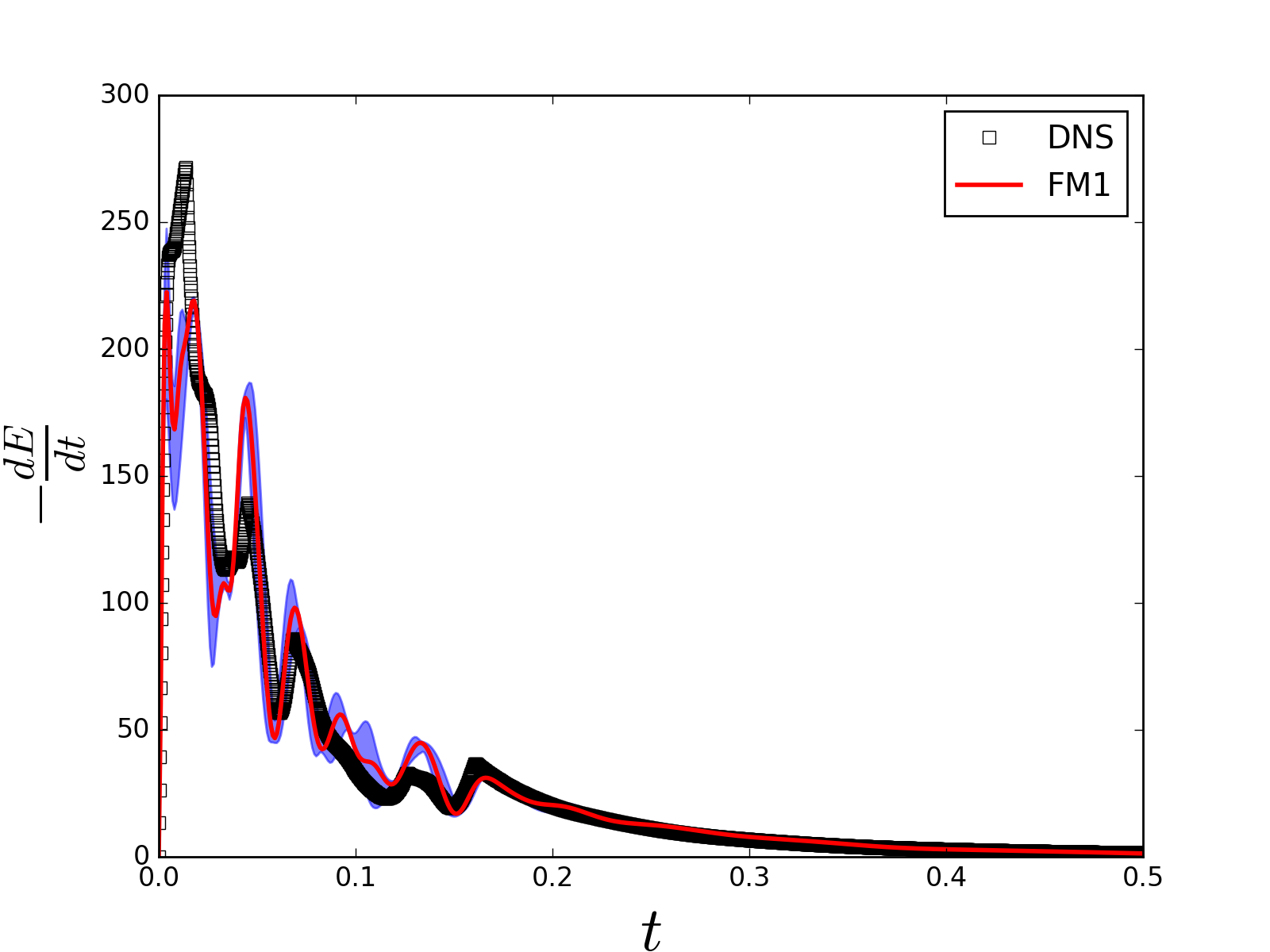}
  \includegraphics[trim={0.05cm 0 1.75cm 1cm},clip,width=.32\linewidth]{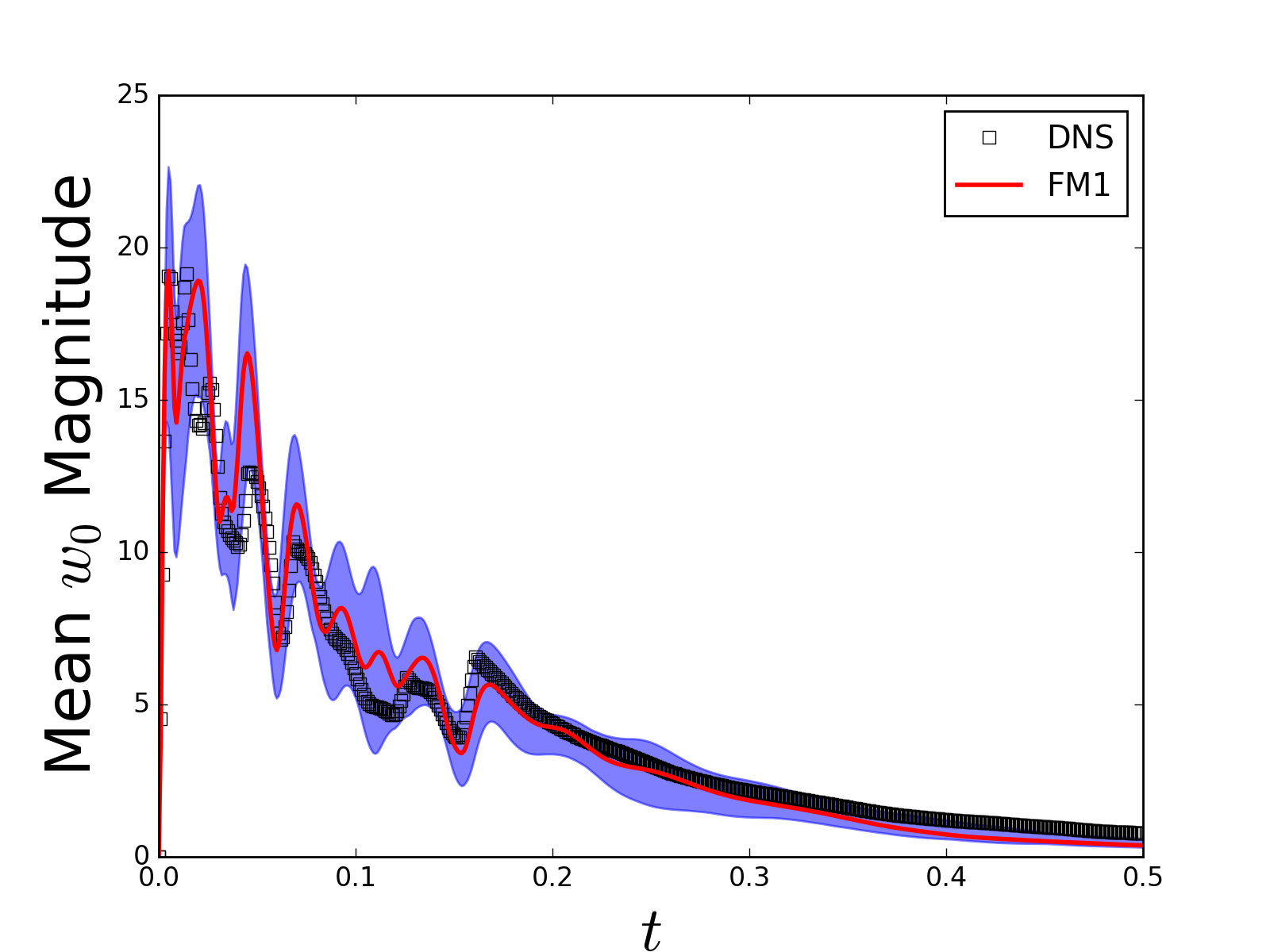}
  \caption{Probability distributions for the VBE case using $k_c=32,\nu=5 \times 10^{-4}$, and $U_0^* = 1.$}
  \label{fig:burg_post2}
\end{subfigure}
\caption{ Probability distributions for the VBE obtained through Monte Carlo sampling. The solid lines indicate the mean of the distribution and the shaded regions are the 95\% confidence intervals.} 
\label{fig:burg_distribution}
\end{figure}

By assuming $\tau_0$ to be stochastic, the results of the Mori-Zwanzig models become random variables. The probability distributions of these variables are determined by Monte Carlo sampling. For each sample, the time constant $\tau_0$ is drawn from the modeled probability distribution and the viscous Burgers equation is evolved from $t=0$ to $t=2$. One-thousand samples are used. The resulting distributions for several of the cases presented earlier are shown in Figure~\ref{fig:burg_distribution}. The probability distributions of integrated first order quantities, such as the total energy and total energy dissipation rate, are seen to be well concentrated around their mean value. The relatively low variance of these distributions shows that concern over selecting the exact optimal time scale $\tau_0$ is not justified for these cases. The variance of the predicted sub-grid content $\mathbf{w}^{(0)}$ is slightly larger than that of the integrated quantities, which can be attributed to the fact that the sub-grid predictions are more dependent on the time scale. The variance of the distributions is still reasonable and the mean values remain accurate. 

\section{A Note on the Physical Space Equivalent of the Finite Memory Models}\label{sec:appendix2}
A physical space equivalent to the finite memory models is derived for the triply periodic case. Define a filtering operation $\overline{G}$ that removes content for $k>k_c$; i.e. a traditional sharp spectral cutoff filter. For the zero-variance projection used in this work, one can write
\begin{equation}\label{eq:physical1}
e^{t\mathcal{L}} \mathcal{P}\mathcal{L} u_i (\mathbf{k},0)=  e^{t\mathcal{L}} \mathcal{P}\mathcal{L} u_i (\mathbf{k},0) -  \overline{ e^{t\mathcal{L}} \mathcal{P}\mathcal{L} u_i (\mathbf{k},0) } \qquad k \in G.
\end{equation}
Evaluating Eq.~\ref{eq:physical1} for the Navier-Stokes equations yields
$$e^{t\mathcal{L}} \mathcal{P}\mathcal{L} u_i (\mathbf{k},0) =  -   A_{im} i k_j \bigg( \sum_{\substack{ \mathbf{p} + \mathbf{q} = \mathbf{k} \\ \mathbf{p,q} \in F} } {u}_{j}(\mathbf{p},t) {u}_m(\mathbf{p},t) -  \overline{ \sum_{\substack{ \mathbf{p} + \mathbf{q} = \mathbf{k} \\ \mathbf{p,q} \in F} } {u}_{j}(\mathbf{p},t) {u}_m(\mathbf{p},t) }\bigg)   \qquad k \in F.$$
The term inside parenthesis on the RHS is the (negative) Leonard stress when the test filter is applied at the same scale as the true filter,
$$e^{t\mathcal{L}} \mathcal{P}\mathcal{L} u_i (\mathbf{k},0) =     A_{im} i k_j L_{jm}.$$
The finite memory model can thus be written as
\begin{multline} \frac{d}{dt} \bigg(-A_{im}  \imath k_j \tau_{jm} \bigg) = - \frac{2}{\tau_0} \bigg(-A_{im}  \imath k_j \tau_{jm} \bigg) + \\ 
2 \bigg( - A_{im} i k_j \sum_{\substack{ \mathbf{p} + \mathbf{q} = \mathbf{k} \\ \mathbf{p,q} \in F} } u_j (\mathbf{p},t) A_{im} \imath k_j L_{jm} (\mathbf{q},t)  - A_{im} i k_j \sum_{\substack{ \mathbf{p} + \mathbf{q} = \mathbf{k} \\ \mathbf{p,q} \in F} } u_m (\mathbf{p},t) A_{jm} \imath k_l L_{ln} (\mathbf{q},t)\bigg). \end{multline}
Next, we note that the projection tensor $A_{im}$ includes the effects of pressure. The physical space model becomes transparent if the effects of pressure on the coarse-graining process is neglected. Neglecting pressure effects simplifies $A_{im}$ to $\delta_{im}$ and one obtains 
\begin{multline}\label{eq:FM_spectral}
\frac{d}{dt} \bigg( -\imath k_j \tau_{ij} \bigg) = - \frac{2}{\tau_0} \bigg(- \imath k_j \tau_{ij} \bigg) + \\ 
2 \bigg( - i k_j \sum_{\substack{ \mathbf{p} + \mathbf{q} = \mathbf{k} \\ \mathbf{p,q} \in F} } u_j (\mathbf{p},t)  \imath k_m L_{im} (\mathbf{q},t)  -  \imath k_j \sum_{\substack{ \mathbf{p} + \mathbf{q} = \mathbf{k} \\ \mathbf{p,q} \in F} } u_i (\mathbf{p},t)  \imath k_m L_{mj} (\mathbf{q},t)\bigg). 
\end{multline}
Taking the inverse Fourier transform yields
\begin{equation}\label{eq:FM_physical}
\frac{\partial}{\partial t} \bigg( \frac{\partial}{\partial x_j}\tau_{ij} \bigg)= -\frac{2}{\tau_0} \bigg(\frac{\partial}{\partial x_j} \tau_{ij}\bigg) + 2 \bigg[ \frac{\partial}{\partial x_j} \bigg( u_j \frac{\partial}{\partial x_m} L_{im} + u_i \frac{\partial}{\partial x_m}L_{jm} \bigg) \bigg],
\end{equation}
where $\tau_{ij}, u,$ and $L_{ij}$ in Eq.~\ref{eq:FM_physical} are the Fourier transforms of their counterparts in Eq.~\ref{eq:FM_spectral}. Interestingly, Eq.~\ref{eq:FM_physical} is not a transport equation. No traditional convection term for $\partial_j \tau_{ij}$ is present. Another interesting observation is that Eq.~\ref{eq:FM_physical} bears a qualitative resemblence to the relaxation equations in the Lagrangian dynamic model~\cite{LagrangianDynamic}. The Lagrangian dynamic model follows the temporal trajectories of fluid particles to determine the Smagorinsky constant. When an exponential weighting function is used, the Lagrangian dynamic model contains a memory time scale that appears in a similar fashion to that which appears in the finite memory models. The construction of physical space M-Z inspired models will be a topic of future research. 
\end{appendices}

\section*{References}
\bibliographystyle{aiaa}

\end{document}